\documentclass[apj]{emulateapj}
\def\gtorder{\mathrel{\raise.3ex\hbox{$>$}\mkern-14mu
                \lower0.6ex\hbox{$\sim$}}}
\def\ltorder{\mathrel{\raise.3ex\hbox{$<$}\mkern-14mu
                \lower0.6ex\hbox{$\sim$}}}

\shorttitle{Detectable Precursors of WHIM Shock}
\shortauthors{O. Gnat}
\slugcomment{}
\begin{document}
\title{Partially Cooled Shocks: Detectable Precursors in the Warm/Hot Intergalactic Medium}
\vspace{1cm}
\author{Orly Gnat\altaffilmark{1,2}}
\altaffiltext{1}{Theoretical Astrophysics, California Institute of Technology, 
        MC 350-17, Pasadena, CA 91125, USA.}
\altaffiltext{2}{Chandra Fellow}
\email{orlyg@tapir.caltech.edu}

\begin{abstract}
I present computations of the integrated column densities produced in the
post-shock cooling layers and in the radiative precursors of partially-cooled fast
shocks as a function of the shock age.
The results are applicable to the shock-heated warm/hot intergalactic medium
which is expected to be a major baryonic reservoir, and contain a large fraction of 
the so-called "missing baryons". My computations indicate that readily observable
amounts of intermediate and high ions, such as \ion{C}{4}, \ion{N}{5}, and
\ion{O}{6} are created in the precursors of young shocks, for which the
shocked gas remains hot and difficult to observe. I suggest that such precursors
may provide a way to identify and estimate the "missing" baryonic mass associated
with the shocks. The absorption-line signatures predicted here may be used to
construct ion-ratio diagrams, which will serve as diagnostics for the photoionized
gas in the precursors. In my numerical models, the time-evolution of the shock
structure, self-radiation, and associated metal-ion column densities are computed
by a series of quasi-static models, each appropriate for a different shock age.
The shock code used in this work calculates the nonequilibrium ionization and
cooling, follows the radiative transfer of the shock self-radiation through the
post-shock cooling layers, takes into account the resulting photoionization and
heating rates, follows the dynamics of the cooling gas, and self-consistently 
computes the photoionization states in the precursor gas. I present a complete 
set of the age-dependent post-shock and precursor columns for all ionization 
states of  the elements H, He, C, N, O, Ne, Mg, Si, S, and Fe, as functions of 
the shock velocity, gas metallicity, and magnetic field. I present my numerical
results in convenient online tables.
\end{abstract}


\section{Introduction}
\label{introduction}

It is a remarkable fact the in the present day universe about a half of the baryonic
matter is "missing", in that it has not yet been accounted for observationally. 
Hydrodynamic simulations of structure formation suggest that these baryons may 
reside in a "warm-hot intergalactic medium" (WHIM), with temperatures in the range
$10^5-10^7$~K (e.g.~Cen \& Ostriker~1999; Dav{\'e} et al.~2001; Bertone et al.~2008). 
The WHIM is produced
by the shock waves that occur as gas falls from the diffuse intergalactic medium 
into the dense regions where galaxies form, and is expected to remain hot because
of the low densities and high temperatures involved.

Recent observations have confirmed the existence of a hot, $10^5-10^7$~K gas, both 
in the local Universe (e.g.~Fang et al.~2006), and in more distant environments 
(e.g.~Tripp et al.~2007; Savage et al.~2005; Buote et al.~2009; Narayanan et al.~2009;
Fang et al.~2010; Danforth et al.~2010). Important ions for its detection 
include \ion{O}{6}, \ion{O}{7}, \ion{O}{8}, \ion{Ne}{8}, and \ion{Ne}{9}
(see also Bertone et al.~2010a, 2010b; Ursino et al.~2010).
In some cases, lower ionization species are also associated with the warm/hot gas. 
For example, Savage et al.~(2005) and Narayanan et al.~(2009) reported on 
absorption-line systems at a redshift of $\sim0.2-0.3$, containing \ion{C}{3}, \ion{O}{3},
\ion{O}{4}, \ion{O}{6}, \ion{N}{3}, \ion{Si}{3}, \ion{Si}{4}, and \ion{Ne}{8}.
They invoked an equilibrium two-phase model, in which most of the ions are created 
in a warm ($\sim2\times10^4$~K) photoionized cloud, while \ion{O}{6} and \ion{Ne}{8}
are created in a hotter ($\sim5\times10^5$~K), collisional phase. The inferred 
temperature for the collisional phase is within the range of temperatures where 
departures from equilibrium ionization are expected to take place 
(e.g.~Gnat \& Sternberg~2007).
While observations have confirmed the existence of warm/hot clouds, it remains
to be determined what fraction of baryons they harbor, and whether they 
confirm the theoretical predictions regarding the shock heated WHIM (Furlanetto et al.~2005).

In Gnat \& Sternberg (2009; hereafter GS09) we computed the metal-absorption column
densities in steady-state fast radiative shocks. We estimated the metal-ion column 
densities in the post-shock cooling layers, under the assumption of steady-state,
completely-cooled shocks (e.g.~Cox~1972; Dopita~1976, 1977; Raymond~1979;
Shull \& McKee~1979; Daltabuit, MacAlpine \& Cox~1978; Binette et al.~1985; 
Shapiro \& Kang~1992; Dopita \& Sutherland~1995, 1996; Allen et al.~2008). 
The steady-state assumption is only valid for shocks
that exist over a time-scale that is longer than their cooling times. In younger
shocks that have not yet cooled completely, the shock structure, self-radiation, and
resulting column densities are time-dependent. For the cosmological shocks that 
produce the WHIM, the steady-state assumption is often not valid.

In this paper, I reexamine the absorption-line signatures of fast astrophysical
shock, but relax the assumption that they are fully-cooled and in steady-state.
I explicitly consider the properties of partially cooled shocks as a function of
shock age. I compute the integrated column densities produced in the post-shock
cooling layers and in the radiative precursors of such partially-cooled fast shocks.
The time-evolution of the shock structure, self-radiation, and associated metal-ion
column densities are evaluated by a series of quasi-static models, each appropriate 
for a different shock age.

I use the shock code presented in GS09
that computes the nonequilibrium ionization and cooling, follows the
radiative transfer of the shock self-radiation through the post-shock cooling
layers, takes into account the resulting photoionization and heating rates, follows
the one-dimensional dynamics of the cooling gas, and self-consistently computes
the photoionization states in the precursor gas.

As in GS09, I focus on fast shocks, with velocities of $600$ and $2000$~km~s$^{-1}$,
corresponding to initial temperatures of $5\times10^6$ and $5\times10^7$~K,
and I present a complete set of results for metallicities $Z$ ranging from
$10^{-3}$ to twice the solar abundance of heavy elements. Investigating how the
absorption line signatures of fast astrophysical shocks depend on metallicity
is particularly interesting in the context of the WHIM, as parts of the WHIM
are expected to have significant subsolar metallicities (e.g.~Cen \& Ostriker~2006).

I consider shocks in which the magnetic field is negligible ($B=0$) so that
cooling occurs at approximately constant pressure, and shocks in which the 
magnetic-pressure dominates the pressures everywhere ($B / \sqrt{\rho} \gg v_s$), 
and the density remains constant. 

In the computations presented here I use the same simplifying assumptions
made in GS09. First, I neglect any thermal instabilities that may form
in the post-shock cooling layers, even though such instabilities are known
to occur for shock velocities in excess of $140$~km~s$^{-1}$.
Second, I assume that the electron and ion temperatures are equal in the 
downstream gas, because the electron-ion equipartition times are typically
$<1\%$ of the cooling time (GS09). This may lead to some errors
for very young shocks that exist for shorter time-scales.
Third, I assume that the shocked material is dust-free, or that any initial 
dust penetrating the shock is rapidly destroyed (e.g.~by thermal sputtering)
on a time scale that is much shorter than the cooling time. For a detailed
discussion of these assumptions see Section~1 in GS09.

In addition to the assumptions of GS09, I make one further assumption when
following the time-dependent evolution of the shock structure. I assume that
the shock evolution is quasi-static, so that it can be approximated by a 
series of ``self-contained'' models for the different ages, which are
independent of each other. Since the shock structure, self-radiation and 
associated precursor ionization states do, in fact, evolve continuously over
the lifetime of the shock, this discretizaton does lead to small errors
in the  computed structure of young shocks. However, as I demonstrate in
the following sections, once the shocks exist for $\gtrsim 1\%$ of their 
cooling times, this approximation is generally valid. In addition, the
quasi-static approximation does not significantly affect the computed 
self-radiation or the integrated column densities.

Therefore, the results for very early times (young shocks with ages $<1\%$
of the shock cooling time) suffer from two main limitations, namely the 
uncertain validity of the quasi-static assumption, and an age comparable 
to the electron-ion equipartition time. 
In addition, at least in the cosmological context, it is not very likely
that we will observe systems with lifetimes that are considerably shorter
than a Hubble time. However, for the sake of completeness, and because the
integrated metal-ion column densities are not very sensitive to these assumptions,
I include young shocks in the discussion that follows.

The outline of this paper is as follows.
In Section~\ref{physics}, I describe the physical processes that I consider in
my computations. These include the ionization, dynamics, cooling and heating, 
radiative transfer, and the treatment of the initial ionization states in the 
gas approaching the shock front.
In Section~\ref{structure}, I discuss the shock structure and emitted radiation, 
and investigate how they depend on the shock age, as a function of the controlling
parameters, including the gas metallicity, shock velocity, and magnetic field.
In Section~\ref{example-columns}, I investigate how the integrated post-shock
column densities evolve as a function of shock age. I describe the full
set of post-shock and precursor column densities versus age in 
Sections~\ref{postcolumns}
and~\ref{precolumns}, respectively. I explain how the UV absorption line
signatures of the radiative precursors may dominate those produced in the
downstream gas over a significant fraction of the shock lifetime, and may thus
serve as means for identifying and detecting young fast shocks.
In Section~\ref{diagnostics}, I demonstrate how ion-ratio diagrams may 
serve as diagnostics for gas in the radiative precursors of partially-cooled
fast astrophysical shocks. The detection and identification of precursor
gas may allow us to confirm the existence of the hot, unobservable, shocked 
counterpart, and to infer its associated baryonic content.
I summarize in Section~\ref{summary}, and conclude in~\ref{conclusions}.








\section{Ionization, Cooling, Dynamics, and Radiative Transfer}
\label{physics}

In this section I describe the physical processes that I take into account in
the shock models, including the ionization, dynamics, cooling and heating, 
radiative transfer, and the treatment of the initial ionization states in the 
gas approaching the shock front. I also describe the method used to follow
the evolution of the shock structure and properties.

After passing the shock front, the gas is initially heated to the shock temperature
$T_s\gtrsim10^6$~K, and then cools and recombines, to a degree determined by the shock age.
I follow a gas element as it advances through the post-shock flow.
As in GS09, if the gas cools faster than it recombines, nonequilibrium ionization
becomes significant, and may affect the cooling rate of the gas. I follow the coupled
evolution of the ionization states and cooling efficiencies.

In this work, I use and improve the shock code developed in GS09.
This code calculates the nonequilibrium ionization and cooling, follows the radiative 
transfer of the shock self-radiation through the post-shock cooling layers, takes into 
account the resulting photoionization and heating rates, follows the dynamics of the 
cooling gas, and self-consistently computes the initial ionization states of the
precursor gas.

For the ionization, I consider all ionization states of the elements H, He, C, N, O, Ne, 
Mg, Si, S, and Fe. I include collisional ionization by thermal electrons, photoionization,
Auger ionization, radiative recombination, dielectronic recombination, and neutralization 
and ionization by charge transfer reaction with hydrogen and helium atoms and ions (GS07,
GS09, and references therein). I follow the time-dependent ionization equation for each 
ion individually (see equation~1 in GS09). I assume the elemental abundances reported by 
Asplund et al.~(2005) for the photosphere of the Sun, and the enhanced Ne abundance 
recommended by Drake \& Testa (2005). I list these abundances in Table~\ref{solar}. I
also  assume a primordial helium abundance $A_{\rm He} = 1/12$ (Ballantyne et al.~2000), 
independent of Z.

\begin{deluxetable}{lr}
\tablewidth{0pt}
\tablecaption{Solar Elemental Abundances}
\tablehead{
\colhead{Element} & 
\colhead{Abundance}\\
\colhead{} &
\colhead{(X/H)$_{\odot}$} }
\startdata
Carbon   & $2.45\times10^{-4}$ \\
Nitrogen & $6.03\times10^{-5}$ \\
Oxygen   & $4.57\times10^{-4}$ \\
Neon     & $1.95\times10^{-4}$ \\
Magnesium& $3.39\times10^{-5}$ \\
Silicon  & $3.24\times10^{-5}$ \\
Sulfur   & $1.38\times10^{-5}$ \\
Iron     & $2.82\times10^{-5}$ \\
\enddata
\label{solar}
\end{deluxetable}

For the cooling, I follow the electron cooling efficiency, $\Lambda(T,x_i,Z)$~(erg~s$^{-1}$~cm$^3$),
which depends on the gas temperature, ion fractions, and abundance of heavy elements.
This includes the removal of electron kinetic energy via collisional excitation followed
by prompt line emission, thermal bremsstrahlung, recombination with ions, and collisional
ionizations (see GS07; based on the cooling functions included in Cloudy ver. 07.02.00, 
Ferland et al.~1998). I also follow the heating rate $\Gamma(x_i,Z,J_\nu)$~(erg~s$^{-1}$),
due to the absorption of the shock self-radiation $J_\nu$, by gas further downstream.
The net local cooling rate is given by $n_en_{\rm H}\Lambda - n\Gamma$, where $n$ is the total
gas particle density.

For the dynamics, I assume one-dimensional strong shocks, and neglect any thermal 
instabilities that may form. I follow the standard Rankine-Hugoniot conditions (see
equations~3 in GS09), which represent the conservation of mass, momentum, and energy
fluxes in the flow. The assumption of a strong shock implies the familiar jump conditions
(e.g.~Draine \& McKee~1993) which relate the pre-shock, and post-shock physical properties: 
$n_0=4n_{\rm pre}$, and $v_0 = 1/4\;v_s$, where $n_0$ and $v_0$ are the post-shock particle 
density and velocity, $n_{\rm pre}$ is the pre-shock density, and $v_s$ is the shock 
velocity. 

I use the Rankine-Hugoniot conditions to follow the evolution of the gas pressure, density and
velocity along the flow. As in GS09, I consider two limiting cases. First, I consider flows
in which $B=0$ everywhere. Second, I consider shocks in the which magnetic field is 
dynamically dominant, with $B/\sqrt{\rho}\gg v_s$, so that it dominates the pressure 
throughout the flow.

When $B=0$ everywhere, the gas cools at an approximately constant 
pressure, and the final pressure far downstream is $P_\infty = 4/3\;P_0$, where $P_0$ is 
the pressure immediately post-shock (see Section~2.3 in GS09 for details).
In the limit of strong magnetic field ($B/\sqrt{\rho}\gg v_s$), the Rankine-Hugoniot conditions
imply constant density and velocity throughout the flow, so that $\rho=\rho_0$, and $v=v_0$ 
everywhere.

To evaluate the intensity of the radiation along the flow, I follow the radiative transfer
equation. As in GS09, I use a Gaussian quadrature scheme (Chandrasekhar~1960) to evaluate 
the local intensity of radiation as a function of distance from the shock front. 
I follow the specific intensity
$I_\nu(\mu)$ along $10$ downstream directions between $\mu=1$ (parallel to the shock velocity)
and $\mu=0$. The flow is divided into thin slabs. In each slab I use Cloudy to evaluate the 
local absorption ($\alpha_\nu$) and emission ($\epsilon_\nu$) coefficients appropriate for 
the local conditions. The radiation transfer equation can then be followed along the $10$ 
values of $\mu$, to calculate the input intensities for the next slab. Finally the mean 
intensity, $J\nu = \frac{1}{4\pi}\int I_\nu(\mu) d\Omega$, is computed locally and used 
in evaluating  the local heating and photoionization rates.

The shock self-radiation also determines the ionization states in the radiative precursor, which
in turn, affects the initial evolution of the post-shock gas. For shock velocities in excess
of $175$~km~s$^{-1}$ such as I consider here, a stable radiative precursor will form  (GS09; 
Dopita \& Sutherland~1996).

In GS09, we allowed the shock to evolve until the gas cooled down to a temperature 
$T_{\rm low} = 1000$~K, at which we terminated the computation.
In this work, I construct shock models as a function of the final shock age.
The shock code used in this work does not produce truly time-evolving computations.
Instead, the evolution of the shock properties with age is approximated by a series
of quasi-steady-state models, each assumed to be independent of other ages as described
below.
For each age, I terminate the computation once a Lagrangian gas element
starting at the shock front exists for a duration that equals the desired age.
The shock structure, self-radiation, and the resulting precursor ionization, are
then functions of this age. For each combination of  shock parameters (temperature, metallicity, 
and magnetic field) I construct a series of such ``stand-alone'' self-consistent models for a 
range of ages.

To self-consistently calculate the coupled ionization states in the precursor and
the post-shock structure, iterations are required. In the first iteration, I assume
that the gas starts in collisional ionization equilibrium (CIE) at the shock 
temperature. I then compute the resulting shock model, ending at the appropriate shock 
age, while following the radiative transfer in the downstream direction.

At early times, the shocks remain optically thin throughout. In this case, I assume
that the flux at the termination point, $F_\nu(l_{\rm age})$ equals the flux
entering the radiative precursor.  I use $F_\nu(l_{\rm age})$ to compute the ionization
states of the gas entering the shock.
For older shocks, the gas becomes thick to its own self-radiation at some large distance
from the shock front $l_{\rm thick}$. In this case, I assume that the flux which ionizes 
the precursor equals $F_\nu(l_{\rm thick})$.
In both cases I use the photoionization code Cloudy (ver. 07.02.00) to compute the
equilibrium photoionization states in the steady precursor. These ionization states 
are the initial conditions for the next iteration. This process is repeated until the 
shock self-radiation and resulting initial ionization states converge.
The assumption that the upstream radiation field equals $F_\nu(l_{\rm thick})$ (or 
$F_\nu(l_{\rm age})$) is based on the fact that the post-shock gas is optically thin 
in this region. Nevertheless, geometrical effects as well as scatterings in the downstream
gas, may increase the intensity of the radiation in upstream directions. This assumption 
may thus lead to some underestimate of the intensity of radiation photoionizing the precursor.

In GS09, we found that two iteration were sufficient to ensure the convergence of the
complete (steady-state) shock models. However, for the younger partial shocks considered here,
I find that more iterations are often required to obtain a self-consistent solution.
The numerical scheme used here is identical to the one described in GS09 (Section~2.6),
with the exception of the different termination criterion. To ensure the convergence
of the different models, I have implemented an algorithm that automatically checks
for convergence and initiates the next iteration if necessary. I found that between
$2$ and $19$ iterations are required, depending on the shock parameters and age.

The approximation used here, where the time-dependent structure of the shock
is approximated by a series of quasi-static, ``self-contained'' models,
introduces some inaccuracies. Since the shock structure, self-radiation and 
associated precursor ionization states do, in fact, evolve continuously over
the lifetime of the shock, the discretizaton may lead to some error in the 
resulting structure of young shocks. 
However, as I demonstrate below, once the shocks exist for $\gtrsim 1\%$
of their cooling times, the approximation is generally valid. In addition, this
approximation does not significantly affect the computed self-radiation or 
resulting integrated column densities.

In the discussion that follows, all ages and times are presented in terms of $n_0\times t$,
the post-shock hydrogen density times time. This scheme makes the results nearly independent
of density (but see Section~3.5 in GS09).

\section{Shock Structure, Ionization \& Cooling}
\label{structure}

In this section, I describe the shock structure, and focus on how it depends on the shock age.
I consider two values of shock temperature: $5\times10^6$~K ($v_s\simeq600$~km~s$^{-1}$)
and $5\times10^7$~K ($v_s\simeq1920$~km~s$^{-1}$), and five different values of the
metallicity $Z$, from $10^{-3}$ to $2$ times the metal abundance of the Sun.
For each shock temperature and gas metallicity I study shocks in the $B=0$ and strong-$B$
limits.

\subsection{A strong-$B$ (isochoric), $T_s=5\times10^6$~K, solar metallicity Shock}
\label{structure-example}

\begin{figure}[!h]
\epsscale{1.18}
\plotone{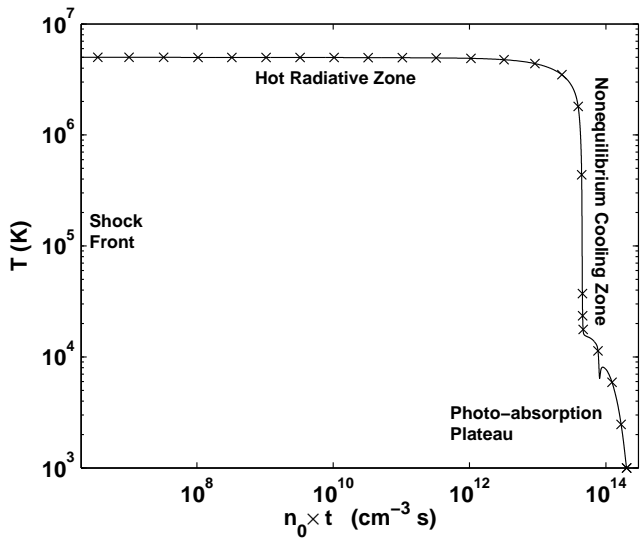}
\caption{Complete shock structure for $T=5\times10^6~K$, $Z=1$~solar
in the strong$-B$ limit. The crosses mark the final ages for which partial
shock models where computed.}
\label{eg-ages-on-complete}
\end{figure}

To illustrate the behavior of the post-shock cooling gas, I first focus on the results
for a strong-$B$ shock with $T_s=5\times10^6$~K, and $Z=1$.
Figure~\ref{eg-ages-on-complete} shows the temperature profile of the post-shock cooling layers
for the complete shock. The crosses along the curve show the times for which partial
shock models were computed.

The complete shock (Figure~\ref{eg-ages-on-complete}) is composed of several zones (GS09). 
The photoionized precursor is heated to the shock temperatures after passing the shock
front. It enters a ``hot radiative'' zone in which cooling is slow. It later passes 
through the ``nonequilibrium cooling'' zone, in which cooling is rapid, and departures 
from equilibrium ionization may occur. After the neutral ion fraction rises, 
photoabsorption heating becomes efficient, and the gas enters a ``photoabsorption plateau'', 
during which the temperature declines more slowly. Finally, after a cooling time, the gas
cool completely. 

This picture is modified for time-dependent shocks that exist over a time-scale shorter 
than their cooling time. I now consider how the shock structure, self-radiation, and 
precursor ionization depend on the shock age.

The initial ionization states of the gas that enters the shock are set by photoionization 
equilibrium of the precursor gas with the shock self-radiation. This radiation field 
builds up as the shock forms. Initially, a young shock emits faint radiation, and the
precursor gas remains largely neutral. As time passes, the shock self-radiation builds up
until the final complete radiation field is obtained after a cooling time.
Even then, the photoionization equilibrium temperature in the radiative precursor is
significantly lower than the shock temperature, and the gas remains underionized compared 
to CIE at $T_s$.

\begin{figure}[!h]
\epsscale{1.18}
\plotone{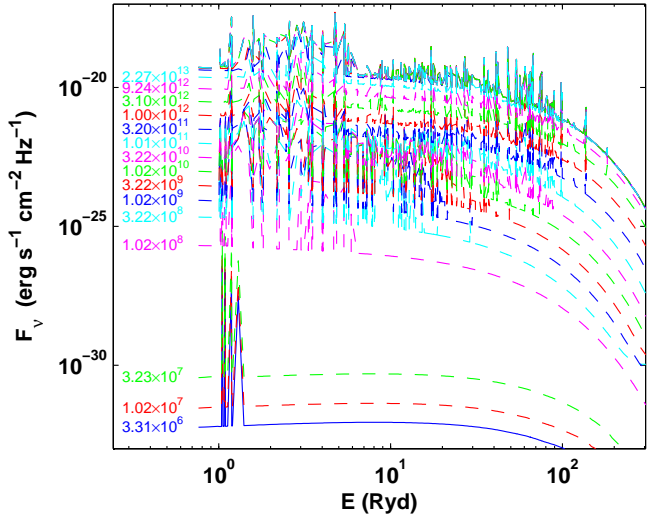}
\caption{Emitted flux by partial shocks with $T=5\times10^6~K$, $Z=1$~solar
in the strong$-B$ limit. Different shock ages ($n_0t$) are shown
by different curves. For ages lower than $2\times10^{13}$~cm$^{-3}$~s the age
is indicated next to the appropriate curve. The curves showing the emitted radiation
for older shocks closely overlaps.}
\label{rad-fields}
\end{figure}

Figure~\ref{rad-fields} shows the radiation flux emitted by a $5\times10^6$~K,
solar metallicity shock at various ages.
Some age-labels are indicated near the spectra for guidance.
Initially, the precursor gas is almost entirely neutral, making electron impact 
excitations and thermal bremsstrahlung emission inefficient.
The three bottom curves in Figure~\ref{rad-fields} show this initial stage, for which
the precursor electron fraction is small ($n_0t < 10^8$~cm$^{-3}$~s).

After the radiation flux builds up as the shock develops, the ionized fraction in
the precursor becomes significant. The various cooling and emission processes are then
much more efficient, and the radiation field rises considerably.
The radiation field continues to grow with shock age, because the
depth of the emitting region increase with time ($10^8 < n_0t < 3\times 10^{13}$~cm$^{-3}$~s).
Finally, as the post-shock cooling layers become thick to the shock self-radiation, the
radiation field entering the precursor stabilizes, and remains constant at later times. 
Figure~\ref{rad-fields} indeed shows that the radiation fields for older shocks nearly
overlap (and labels are therefore not shown for $n_0t > 3\times 10^{13}$~cm$^{-3}$~s).

\begin{figure}[!h]
\epsscale{1.18}
\plotone{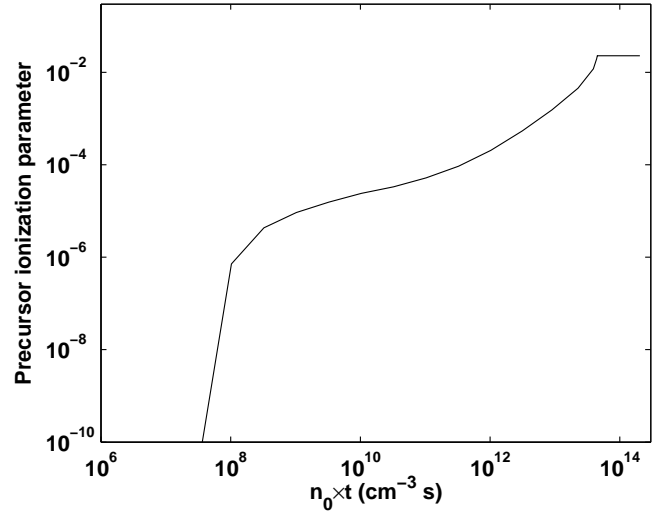}
\caption{Ionization parameter in the radiative precursor as a function of shock age
for partial shocks with $T=5\times10^6~K$, $Z=1$~solar in the strong$-B$ limit.}
\label{U}
\end{figure}

\begin{figure}[!h]
\epsscale{1.18}
\plotone{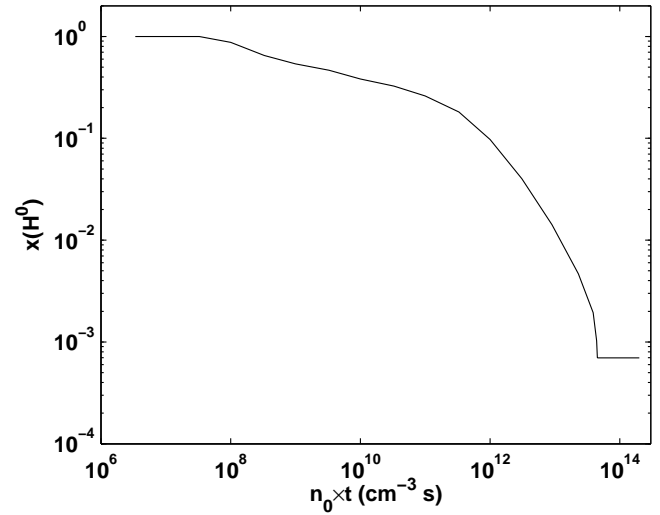}
\caption{Neutral fraction in gas entering shock, as a function of shock age
for partial shocks with $T=5\times10^6~K$, $Z=1$~solar in the strong$-B$ limit.}
\label{x0}
\end{figure}

In Figure~\ref{U}, I show the ionization parameter that the shock self-radiation
produces in the radiative precursor as a function of shock age. 
Initially, the ionization parameter is very low, but it increases with shock age,
until finally stabilizing at a value appropriate for the complete shock.
Figure~\ref{x0} shows the corresponding neutral fraction in the material entering 
the shock.
Initially ($n_0t \lesssim 10^8$~cm$^{-3}$~s), the precursor gas is almost entirely neutral.
The lack of free electrons leads to the initial inefficient radiation demonstrated in 
Figure~\ref{rad-fields}.
At later times, the neutral fraction begins to decrease. First, both free electrons
and neutral hydrogen coexist in the precursor gas, making initial cooling and emissivity
in the post shock layers very efficient. As the depth of the radiating layers increases,
hydrogen becomes mostly ionized, and the cooling efficiency of the post-shock
layers drops again. It always remains higher that the CIE cooling efficiency due to
underionized metal species that are penetrating from the radiative precursor.

\begin{figure}[!h]
\epsscale{1.18}
\plotone{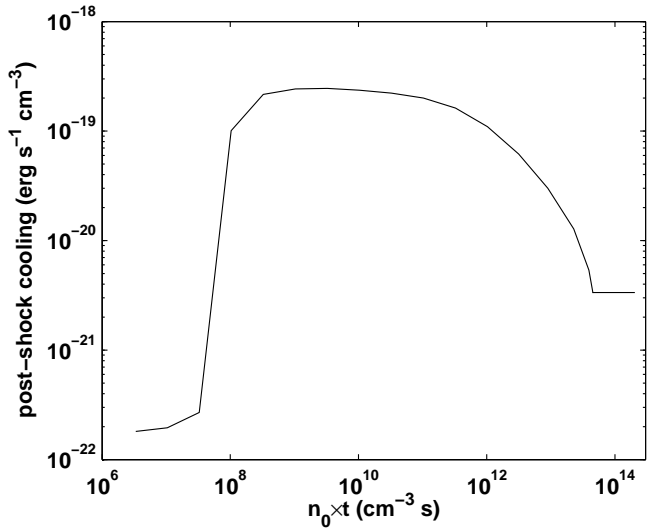}
\caption{The initial cooling rate (erg~s$^{-1}$~cm$^{-3}$) as a function of shock age,
for partial shocks with $T=5\times10^6~K$, $Z=1$~solar in the strong$-B$ limit.}
\label{init_cool}
\end{figure}

The initial post-shock cooling rates as a function of age are displayed in 
Figure~\ref{init_cool}. This figure confirms that initially the lack of free
electrons suppresses the cooling rate. The cooling rises as the electron fraction
grows, until a maximal value is obtained when both neutral hydrogen atoms and
free electrons are abundant in the gas entering the shock. At later times,
as the gas become highly ionized, the abundance of efficient coolant decreases
and the cooling rate declines again.

\begin{figure}[!h]
\epsscale{1.18}
\plotone{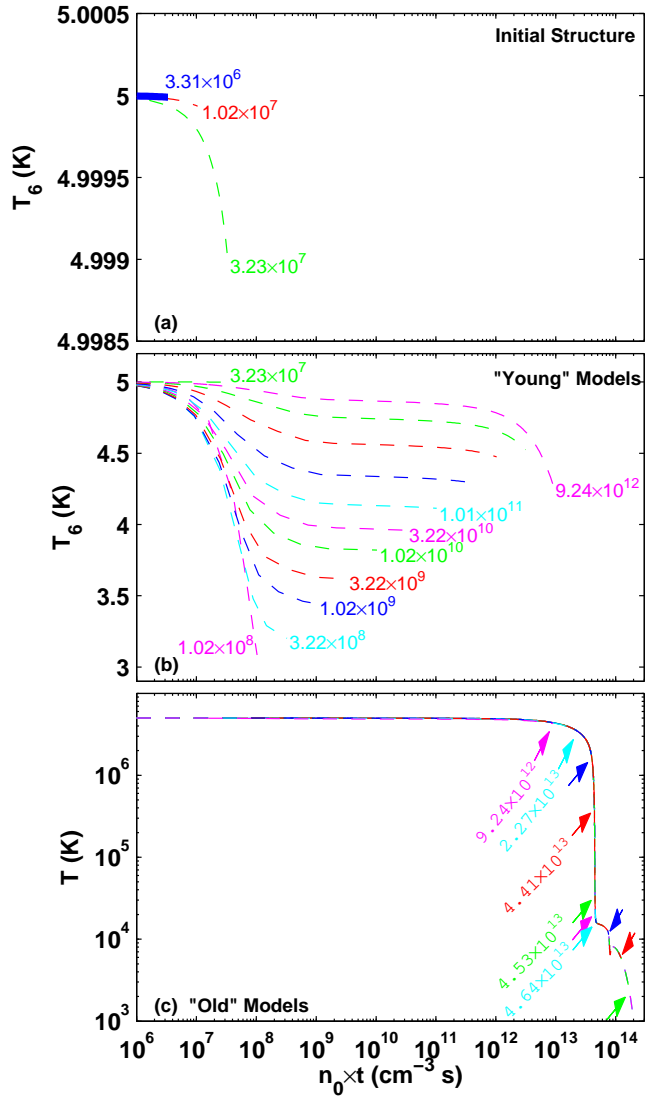}
\caption{Temperature as a function of time for partial shocks with various ages
for $T=5\times10^6~K$, $Z=1$~solar in the strong$-B$ limit.
(a) Initial evolution. Following the shock formation, the incoming electron electron
fraction is small, and electron impact excitations are inefficient. Cooling
becomes more rapid as the shock evolves and the incoming neutral fraction grows.
(b) ``Young'' shocks. In these shocks both neutral hydrogen and electrons are abundant
at the shock front. Younger shocks cool more efficiently due to hydrogen Ly$\alpha$ 
cooling. Older shocks contain a smaller neutral fraction.
(c) ``Old'' shocks. The incoming gas is ionized. These shocks follow a part
of the trajectory of the complete shock.}
\label{Tts}
\end{figure}

The impact that this has on the shock structure is demonstrated in Figure~\ref{Tts}.
The upper panel shows the initial shock structure following the shock formation.
As mentioned above, initially the free electron fraction is negligible, and cooling is 
inefficient. With time, the electron fraction grows, and cooling becomes more efficient.
For example, panel~(a) shows that the initial cooling at an age of 
$n_0t=3.23\times10^7$~cm$^{-3}$~s
is more rapid than the cooling at $n_0t=1.02\times10^7$~cm$^{-3}$~s.

\begin{figure}[!h]
\epsscale{1.18}
\plotone{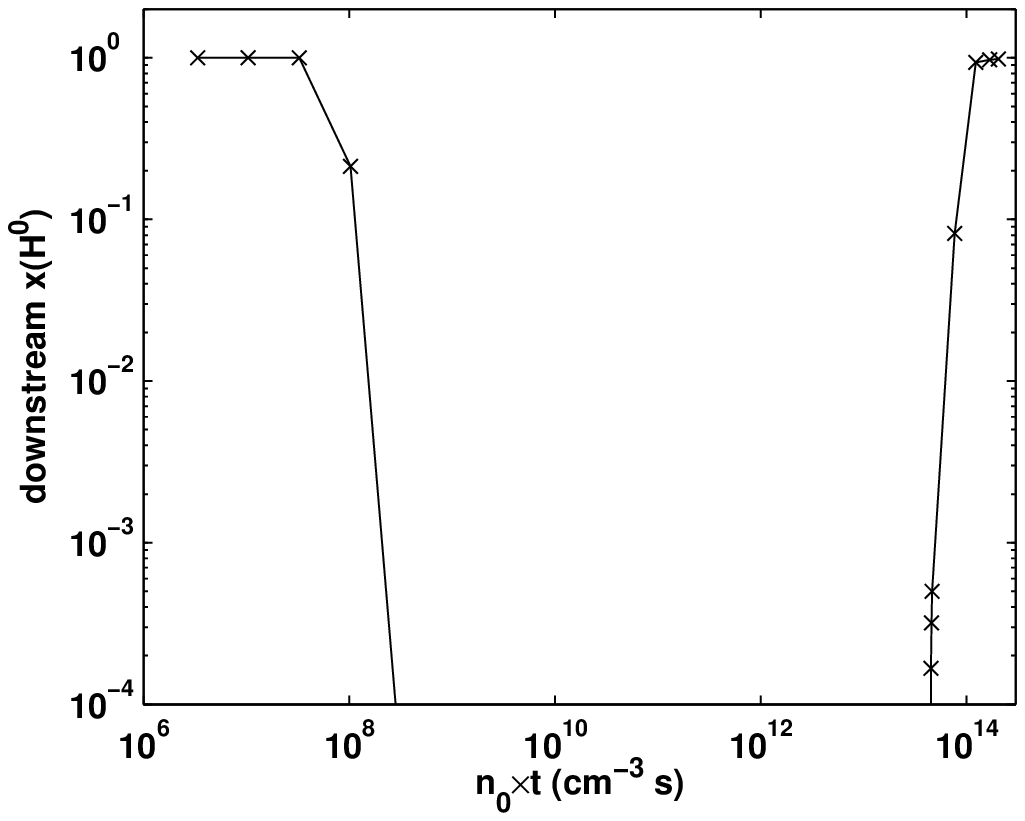}
\caption{The downstream neutral fraction as function of shock age,
for partial shocks with $T=5\times10^6~K$, $Z=1$~solar in the strong$-B$ limit.}
\label{xHIend}
\end{figure}

For these early ages, the gas remains neutral
in the post-shock cooling layers, as the ionization time is longer than the
shock age. This is demonstrated in Figure~\ref{xHIend}, in which I show the
{\it final} neutral hydrogen fraction, obtained at the maximum distance downstream 
from the shock front, as a function of the shock age.
The figure shows that for ages $\lesssim 10^8$, the gas remains partially neutral
for the duration of the shock age. At later times, two effects act to increase the
final ionization. First, the shock age becomes longer than the ionization time-scale, 
and second the initial precursor neutral fraction declines.
It is important to note that while the gas is being ionized,
the self-radiation propagating upstream in the post-shock layers, which is not
included in this computation,
may affect the shock structure and ionization states. This only affects very
young shocks (panel~(a) in Figure~\ref{Tts}), and therefore has little impact
of the results presented in this paper.

Returning now to Figure~\ref{Tts}.
Once a significant electron fraction is established, the cooling depends on the availability
of coolants in the precursor gas. This phase is shown by the middle panel. For a shock
age of $10^8$~cm$^{-3}$~s, neutral hydrogen is abundant, and the initial cooling is very 
efficient, both due to collisional ionization and line cooling. This shock cools to a 
temperature of $\sim3\times10^6$~K within its life time.
At later times, the shock self-radiation increases the ionization of the radiative precursor, 
and the initial post-shock cooling efficiency diminishes. 
The shortcomings of the quasi-static approximation are demonstrated in this panel.
With the assumption that the age dependent structure may be approximated by a series of
``self-contained'' steady-states models, younger shocks cool to lower final temperatures
than their older counterparts. For example, at an age of $9.2\times10^{12}$~~cm$^{-3}$~s, 
the shock only cools down to $\sim4\times10^6$~K. This inaccuracy in the temperature
profile computations diminishes for older shocks, where the self-radiation and associated
precursor ionization stabilize and are no longer a strong function of shock age.
Because the self-radiation depends more on the total age of the shock and on the
metal abundance (rather than on the details of the temperature profile), the computed
self-radiation is not significantly affected by the quasi-static approximation.
As I discuss below, the predicted column densities and resulting absorption line signatures
are also independent of the quasi-static approximation.

After $\sim10^{13}$~cm$^{-3}$~s, the shock self-radiation approaches its maximal intensity, 
and at later 
times any additional contribution to the shock self-radiation is negligible. The initial
ionization states of the gas in the radiative precursor then depend only weakly on the 
shock age, and no longer affect the dynamical evolution of the shock. This is shown in 
the bottom panel. The temperature profiles of all the shocks with ages larger than 
$9.24\times10^{12}$~~cm$^{-3}$~s overlap at early time. The arrows mark the ending 
points of partial shocks of various ages, and some labels are indicated near the arrows
for guidance.

\subsection{Gas Metallicity}

The gas metallicity affects the shock structure in two major ways (GS09).
First, the metal content affects the cooling efficiency. For high-metallicity shocks, 
cooling is dominated by metal-line emission. For $10^4\lesssim T\lesssim 10^6$~K and 
$Z\gtrsim0.1$, permitted transitions dominate the gas emission, and the cooling is roughly 
proportional to $Z$. At lower gas metallicities, cooling is dominated by hydrogen and 
helium. Below $10^4$~K, cooling is dominated by metal fine-structure transitions
even at low metallicities. The cooling times are then always roughly proportional 
to $Z$. Higher metallicity shocks therefore evolve more rapidly. The different cooling 
times also set the degree of non-equilibrium ionization in the cooling zone. Departures
from equilibrium ionization are therefore larger at higher metallicities.

Second, while the total energy emitted by the complete shock
is independent of gas metallicity ($F\propto n_0v_s^3$) and is set by the 
input energy flux into the flow, the spectral energy distribution of the shock
self-radiation is a strong function of gas metallicity. For low-$Z$, the radiation 
is dominated by thermal bremsstrahlung emission. The resulting spectrum is therefore
a smooth bremsstrahlung continuum, with very little features. At higher metallicities, 
the contribution of metal line emission increases. This has a major effect on the shape of the
emitted spectrum. For example, Figure~\ref{rad-fields} demonstrates that the emission
of a $T_s=5\times10^6$~K, solar metallicity shock is dominated by numerous
metal emission lines. This affects both the downstream ion fractions (for ``old'' enough shock)
and the ionization states in the radiative precursor.

\begin{figure}[!h]
\epsscale{1.18}
\plotone{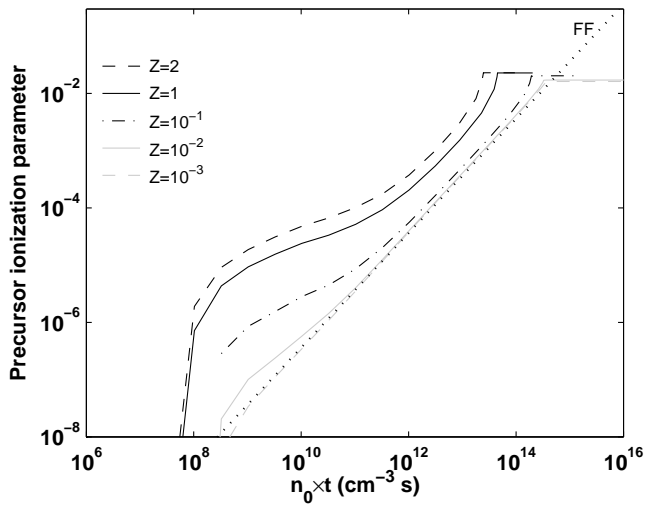}
\caption{Ionization parameter in the radiative precursor as a function of shock age
for different gas metallicities, between $Z=2$ and $Z=10^{-3}$ times solar.
The dotted curve shows the ionization parameter due to thermal bremsstrahlung emission
at the shock temperature, $T_s=5\times10^6$~K. The low metallicity results ($Z=10^{-3}$)
overlap with the Bremsstrahlung predictions, but higher metallicity shocks produce
larger ionization parameters due to line-emission.}
\label{UvsZ}
\end{figure}

Figure~\ref{UvsZ} shows how the changing spectral energy distribution affects
the ionization parameter in the precursor gas. The different curves in Figure~\ref{UvsZ}
show the precursor ionization parameter for different gas metallicities, as indicated
by the labels, as a function of shock age.
At low metallicities (grey curves) the ionization parameter is a simple power
law in the shock age. This can be simply understood in terms of ionization
by a thermal bremsstrahlung continuum. The thermal bremsstrahlung emissivity 
at the shock temperature (erg~s$^{-1}$~cm$^{-3}$) is given by,
\begin{equation}
\varepsilon_{\rm FF}(\nu) = 6.8\times10^{-38} T_s^{-0.5} e^{h\nu/k_{\rm B}T_s}\;.
\end{equation}
Assuming that a layer with depth $v_{\rm ps}\times t$ radiates with the above emissivity,
the ionization parameter can be approximated by,
\begin{equation}
U_{\rm FF} = \int{\frac{\varepsilon_{\rm FF}(\nu) v_{\rm ps}}{h\nu}}\; \frac{1}{c}\; t\; d\nu\;,
\end{equation}
where $v_{\rm ps}$ is the initial post-shock velocity.
This simple approximation is shown by the dotted line is Figure~\ref{UvsZ} (denoted ``FF'').

The results for $Z=10^{-3}$ overlap with the predicted bremsstrahlung ionization 
parameter. However, higher metallicity shocks produce larger ionization parameters,
as a large fraction of their initial energy flux is emitted via line emission
at ionizing energies, and a smaller fraction is emitted as high-energy free-free photons.
For example, the solar metallicity curve (solid) is higher by $0.5$ to $2$ dex
compared with the bremsstrahlung predictions. This implies that the radiative
precursor is more highly ionized at any age for higher metallicity shocks.
The ionized faction, and implied initial cooling rates vary accordingly.

\subsection{Shock temperature}

To study the impact of the shock temperature on the shock structure and evolution,
I compare the results presented in section~\ref{structure-example} for
$T_s=5\times10^6$~K with a model of a $5\times10^7$~K shock.
The overall characteristics of the hotter shock are similar to those discussed
above. However, the shock temperature, and associated velocity, affect the 
initial energy content of the shock. Hotter shocks therefore have longer overall
cooling times.  For example, in the $5\times10^7$~K shock the initial temperature
is $10$ times higher, and the velocity is $10^{1/2}$ times higher than in the
$5\times10^6$~K shock. The total energy flux input is therefore $\sim30$
times larger. This affects the total cooling time of the shock.

The emitted radiation field is also a strong function of the shock temperature.
It is affected by the increased initial temperature.
Above $\sim2\times10^7$~K, bremsstrahlung is the dominant emission process
in CIE even at high metallicities. In complete shocks, the thermal energy
distribution is therefore dominated by a hard free-free continuum.
This continuum extends to higher energies than for $5\times10^6$~K, because 
the exponential cut-off occurs at higher temperatures. The spectrum is also
considerably smoother, because a large fraction of the initial energy flux
is emitted as bremsstrahlung continuum, and the intense line emission that
occurs at lower temperatures as the gas cools only provides a small contribution.

\begin{figure}[!h]
\epsscale{1.18}
\plotone{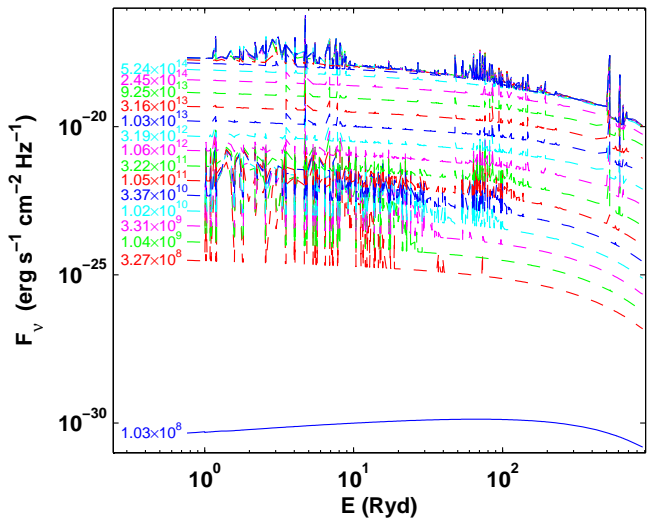}
\caption{Emitted radiation for a $T_s=5\times10^7$~K, $Z=1$ shock in the strong-B limit.
At early times, the precursor gas is significantly underionized, leading to efficient line
emission following the shock. However, as the gas ionization level increases, the
post-shock radiation at $5\times10^7$~K becomes dominated by thermal bremsstrahlung.}
\label{Ts-rad}
\end{figure}

For partial shocks, however, line emission may play a dominant role even
at $T_s=5\times10^7$~K shocks. Figure~\ref{Ts-rad} shows the emitted radiation
as a function of shock age for a $5\times10^7$~K, solar metallicity shock
in the strong-$B$ limit. After the gas becomes partially ionized 
($n_0\times t \ge 3.3\times10^9$~cm$^{-3}$~s) the increased availability of
metal lines coolants produces a significant emission-line UV-bump in the
resulting spectrum. This is possible due to the highly underionized precursor
that obtains at early times, before the full intensity of the shock self-radiation 
has accumulated.
At later times, as the shock self-radiation efficiently ionizes the precursor gas,
the fractional contribution of metal lines decrease, and a smoother spectrum is
produced. For $3.2\times10^{12} < n_0\times t < 5.2\times10^{14}$~cm$^{-3}$~s, the
precursor gas is ionized, but the shocked-gas temperature is everywhere high enough
to impede significant contribution of lines. At later times, at the gas cools in the 
post-shock flow, metal lines again contribute significantly to the emitted spectrum.
The resulting spectrum has a lower line-to-continuum contrast, and extend to higher 
energies than the $5\times10^6$~K shock (c.f. Figure~\ref{rad-fields}).

\begin{figure}[!h]
\epsscale{1.18}
\plotone{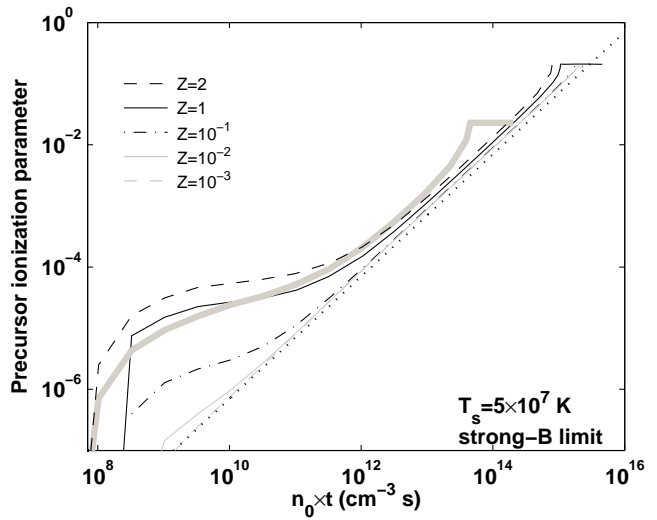}
\caption{The precursor ionization parameter in a $T_s=5\times10^7$~K shock 
in the strong-B limit. The different curves show the ionization parameter for different
gas metallicities as a function of shock age. The dotted curve shows the ionization 
parameter due to thermal bremsstrahlung emission at the shock temperature, $T_s=5\times10^7$~K.
The low metallicity results ($Z=10^{-3}$) overlap with the Bremsstrahlung predictions.
However, at late times even the high-$Z$ cases agree with the bremsstrahlung predictions.
For comparison, the thick gray curve shows the ionization parameter for $T_s=5\times10^6$~K.
}
\label{Ts-U}
\end{figure}

In Figure~\ref{Ts-U}, I show how these spectral variations affect the ionization parameter
in the radiative precursor. The different curves in Figure~\ref{Ts-U} show the precursor 
ionization parameter as a function of shock age for gas metallicities between
$10^{-3}$ and $2$ times solar.
As before, I compare these numerically computed ionization parameters with the 
expectation from a pure bremsstrahlung continuum at the shock temperature (see 
equations~1-2) shown by the dotted line.
The low metallicity results are again consistent with the predictions of the free-free continuum.
However, as opposed to the results presented in Section~\ref{structure-example} for the 
$5\times10^6$~K shock, Figure~\ref{Ts-U} shows that for $T_s=5\times10^7$~K, the late time 
(age $\gtrsim10^{12}$~cm$^{-3}$~s) results for high-$Z$ are also nicely consistent 
with the bremsstrahlung
expectations. As discussed above, this is because at high temperatures, a large faction of
the initial energy is radiated as free-free emission, and the total contribution of 
metal emission lines is smaller. At early times, when underionized gas rich with 
low-ionization species enters the shock front, the line contribution to the spectral energy 
distribution for high metallicity shocks remains significant.

For comparison, the thick gray curve shows the results for a $5\times10^6$~K shock.
The final ionization parameter produced by the hotter shock is considerably
larger, as the higher energy content of the faster shock is radiated away.

\subsection{Magnetic field}

The results presented so far have demonstrated the characteristics of shocks in the
strong-$B$ (isochoric) limit.
Shock for which $B=0$ evolve differently at late times. The most important difference
is that  when $B=0$, the pressure remains nearly constant, and the density increases
as the gas cools. Since the cooling time is inversely proportional to the gas density,
this implies rapidly decreasing cooling times (fast cooling) within the flow.
The overall cooling times are much shorter, and in particular, the evolution below
$10^4$~K is compressed into a very short time interval.

The final evolution of $B=0$ shocks is affected not only by the shorter cooling 
times, but also by the difference in downstream ionization parameter, $\propto F/n$.
For strong-$B$ shock, the density in the flow is constant and the ionization
parameters is $\propto v_s^3$. For $B=0$, the density increases as the gas cools,
and the ionization parameter is therefore $\propto v_s^3n_0/n$, a factor $n_0/n$
smaller. Photoionization in the downstream layers is much less important when $B=0$.
For the nearly isobaric evolution that obtains when $B=0$, $n_0/n\propto T/T_0$.
The ionization parameter and resulting ion fractions are therefore significantly 
affected in the downstream gas after significant cooling has occurred.
However, the differences in the post-shock ionization structure at early ages, 
before much cooling took place, remain small.

A second effect is related to the $PdV$ work that appears in $B=0$ shocks. The work
done on the cooling gas implies that the overall emitted radiation is $5/3$ times 
larger for $B=0$ shocks than for isochoric shocks. Other than this $5/3$ 
scaling in the intensity of the integrated emission, however, the radiation emitted by 
$B=0$ shocks is similar to that produced by shocks in the strong-$B$ limit.
The ionization in the radiative precursor is thus nearly independent of magnetic 
field.

\begin{figure}[!h]
\epsscale{1.18}
\plotone{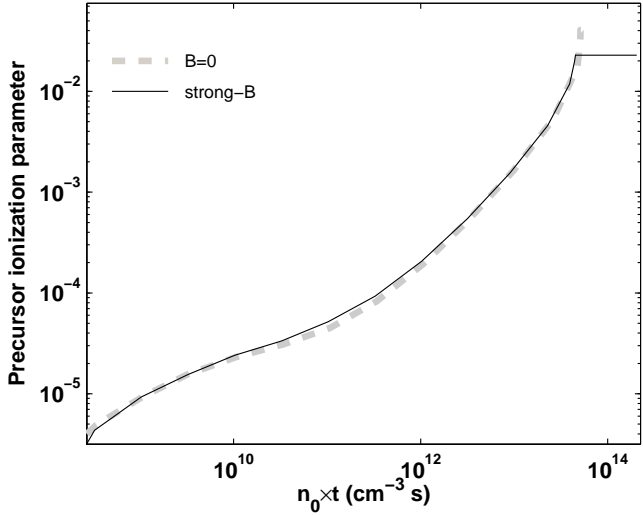}
\caption{A comparison of the precursor ionization parameter for $B=0$ (gray dashed line)
and in the strong-B limit, for shocks with $T_s=5\times10^6$~K and solar metallicity.
The two curves nearly overlap at early times. The final ionization parameter of
the complete shock is $5/3$ times higher when $B=0$, due to $PdV$ work.}
\label{B-U}
\end{figure}

Figure~\ref{B-U} demonstrated both points. It compares the precursor ionization parameter
for $B=0$ (gray dashed curve) and for the strong-$B$ limit (solid curve).
Despite the differences in the dynamical evolution, the two curves overlap at early times.
It is only at late times that differences between the two cases arise.
The final ionization parameter for $B=0$ is $5/3$ times larger than in the strong-$B$
limit, due to contribution of the $PdV$ work.

\section{The Ionization States and Metal Columns in a Strong-B, $5\times10^6$~K, Solar Metallicity Shock}
\label{example-columns}

\begin{figure}[!h]
\epsscale{1.18}
\plotone{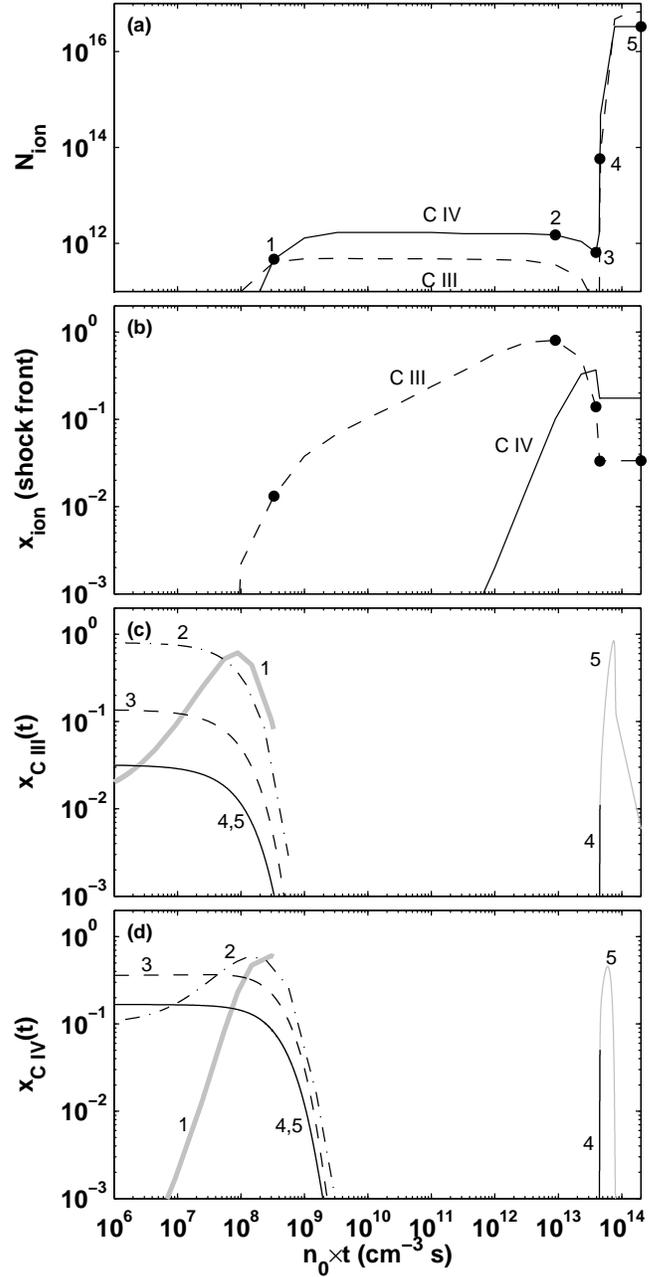}
\caption{(a) Column densities versus shock age for a $T=5\times10^6~K$, $Z=1$~solar 
metallicity shock in the strong$-B$ limit. The dashed curve is for \ion{C}{3} and 
the solid curve for \ion{C}{4}.
(b) Ion fraction at the shock front as a function of shock age.
(c) \ion{C}{3} ion fraction as a function of {\it time} in five shock models.
The final ages of the shocks shown here are indicated by the filled symbols in
the top panel, and marked ``1''-``5''.
The thick gray curve is for a shock age of $3\times10^9$~cm$^{-3}$~s (``1'');
The dash-dotted curve is for $\sim10^{12}$~cm$^{-3}$~s (``2'');
the dashed curve is for $3.9\times10^{13}$~cm$^{-3}$~s (``3'');
the dark solid curve is for $4.5\times10^{13}$~cm$^{-3}$~s (``4'');
and the thin gray curve is for the complete shock (``5'').
The curves for the last two cases overlap at early times.
(d) Same as (c) but for \ion{C}{4}.}
\label{C-cols}
\end{figure}

In this section I present computation of the integrated post-shock metal-ion columns 
produced in a strong-$B$ shock with $T_s=5\times10^5$~K and solar
metallicity. As an example, I discuss the formation of \ion{C}{3} and \ion{C}{4}
behind the shock front. Figure~\ref{C-cols}a shows the \ion{C}{3} and \ion{C}{4} column
densities created in the post-shock cooling layers as a function of shock age.
For ages shorter than $\sim10^8$~cm$^{-3}$~s, \ion{C}{3} and \ion{C}{4} are not created
behind the shock. At this age, there is a rapid increase in the column, which later
stabilizes at a level of $\sim10^{12}$~cm$^{-2}$. The columns then remain nearly constant
for $\sim10^{13}$~cm$^{-3}$~s. After that, the \ion{C}{3} and \ion{C}{4} columns decrease,
before rising again to achieve the levels observed in the complete shock.

To explain this behavior, the evolution of the carbon ion fractions in the post-shock
cooling layers must be understood. I focus on five representative ages, marked
by the filled symbols in panel (a). These representative ages are $3\times10^9$~cm$^{-3}$~s (``1'');
$\sim10^{12}$~cm$^{-3}$~s (``2'');  $3.9\times10^{13}$~cm$^{-3}$~s (``3'');
$4.5\times10^{13}$~cm$^{-3}$~s (``4''); and the complete shock (``5'').

Panel (b) shows the corresponding initial precursor \ion{C}{3} and \ion{C}{4} ion fractions
as a function of shock age. \ion{C}{3} is efficiently produced in the radiative precursor
at times later than $\sim10^8$~cm$^{-3}$~s. Before that, the ionization parameter is smaller,
and lower ionization states of carbon dominate its ion distribution.
\ion{C}{3} reaches the peak of its precursor abundance at a shock age of 
$\sim10^{13}$~cm$^{-3}$~s,
after which it is replaced by \ion{C}{4}. At $\sim4\times10^{13}$~cm$^{-3}$~s
\ion{C}{4} is replaced by even higher stages of ionization.
The five representative times are shown on the \ion{C}{3} abundance curve for comparison.

In the post shock cooling layers, the gas adjusts to CIE at the shock temperature.
At $5\times10^6$~K, the most abundant carbon ion in CIE is fully stripped carbon (C$^{6+}$).
Given sufficient time, the shocked gas will approach the CIE ionization state.
At very early times ($<10^8$~cm$^{-3}$~s) when the radiative precursor is effectively neutral
and the free electron fraction is negligible (see Section~\ref{structure}), the shocked gas cannot
efficiently cool or become ionized. Therefore, despite the initially neutral carbon ionization,
no \ion{C}{3} and \ion{C}{4} are produces in the cooling layers during the shock's short lifetime.
After this time, the post-shock gas does  contain a significant fraction of free electrons,
and the ionization and cooling processes become more rapid. At this stage, the gas can
rapidly adjust to CIE near the shock temperature. The initial carbon ion distribution,
which is dominated by low and intermediate-ions, is rapidly replaces by high-carbon-ions,
until the carbon ion fractions finally establish the appropriate CIE distribution.
Some amounts of \ion{C}{3} and \ion{C}{4} are produced in this rapid
initial ionization stage during which the gas adjust to CIE at the shock temperature.
This is the column density seen in points ``1'' and ``2'' in panel~(a).
Because these initial columns are produced over an ionization time scale which is
significantly shorter than the shock age, they are insensitive to the exact temperature
profile of the post-shock gas, and are not significantly affected by the quasi-static
approximation.

The specific ion distributions are shown in panels~(c) (for \ion{C}{3}) and (d) (for \ion{C}{4});
These panels follow the ion fractions as a function of {\it time} (not age) for the different
shock ages shown in panels~(a) and (b).
For example, curve ``1'' corresponds to a shock age of $3\times10^9$~cm$^{-3}$~s.
Panel~(b) shows that at this age, the initial \ion{C}{3} ion fraction is $\sim10^{-2}$,
and the initial \ion{C}{4} ion fraction is negligible. The precursor carbon ion distribution
is dominated by lower ionization states.
Panel~(c) shows that the \ion{C}{3} abundance initially grows as the
gas get more highly ionized, until it decreases again to be replace by \ion{C}{4} (panel~(d)).
The resulting column densities are shown in panel~(a).

Point ``2'' shows a later age, for which the precursor \ion{C}{3} ion fraction is at its
peak, and the \ion{C}{4} abundance is also significant (panel~b).
Panel~(c) the shows that \ion{C}{3}, which starts at its peak abundance, is rapidly
replaced by \ion{C}{4} (panel~(d)), which in turn is later replaced by higher ionization 
states.
Point ``3'', at an age of $3.9\times10^{13}$~cm$^{-3}$~s, is right at the minimum of
the \ion{C}{3} and \ion{C}{4} column densities. Panel~(b) shows that at this age the radiative
precursor is ionized enough that \ion{C}{4} is the dominant ion, whereas the abundance of
\ion{C}{3} is already reduced. Panel~(c) shows that after starting at an abundance
of $\sim0.1$, \ion{C}{3} is rapidly replaced by the initially dominant \ion{C}{4}. The ion
fraction of \ion{C}{3} is everywhere lower than that which existed at the age shown
by point ``2'', thus explaining the reduction in the column density associate with this
initial post-shock ionization stage.

Finally, in ages ``4'' and ``5'', the shock exists for long enough to allow for the
production of \ion{C}{3} and \ion{C}{4} in the downstream recombination layers of the shock.
Even though the contribution of the initial ionization stage to the total column is
lower than it was at earlier times, the contribution from late-time production of these
ions, which spans a considerably larger path length, causes the rapid and substantial 
increase in the columns seen in panel~(a). Finally (point 5) the total columns approach 
the values obtained in the complete shock.

\begin{figure}[!h]
\epsscale{1.18}
\plotone{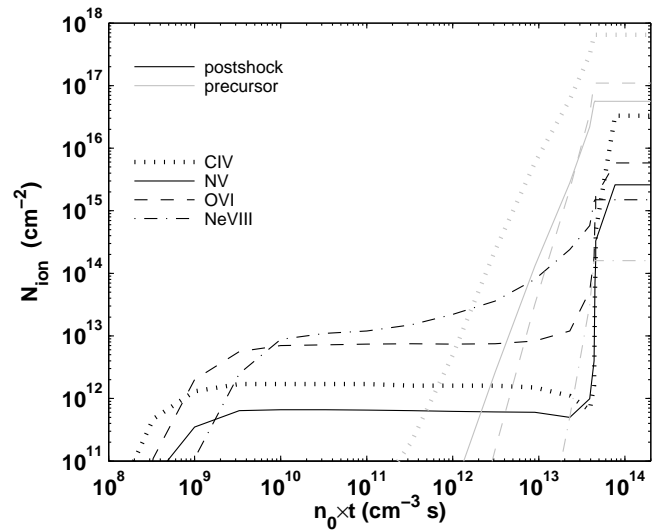}
\caption{Post-shock and precursor column densities (cm$^{-2}$) as a function of
shock age ($n_0\times t$) for \ion{C}{4}, \ion{N}{5}, \ion{O}{6}, and \ion{Ne}{8}.
in a $T=5\times10^6~K$, $Z=1$ shock in the strong$-B$ limit.}
\label{col-ex}
\end{figure}

Figure~\ref{col-ex} shows examples for the column densities produced in partial shocks
with $T_s=5\times10^6$~K, $Z=1$, in the strong-$B$ limit. The dark curves show the columns
densities produced in the post-shock cooling layers as a function of shock age. The dotted
curves corresponds to \ion{C}{4} (and shows the same curve displayed in Figure~\ref{C-cols}).
The other curves are for \ion{N}{5}, \ion{O}{6}, and \ion{Ne}{8}.
Figure~\ref{col-ex} also displays the column densities produced in the radiative precursor
due to the shock self-radiation as a function of shock age. These are shown by the gray curves.
It is clear from Figure~\ref{col-ex} that significant amounts of intermediate/high ions are produced
in the radiative precursor {\it before} similar columns are produced in the 
post-shock cooling layers.
For example, at an age of $2\times10^{12}$~cm$^{-3}$~s, the precursor \ion{C}{4} column density
is $\sim10^{14}$~cm$^{-2}$, whereas the post-shock column at this age is only $\sim10^{12}$~cm$^{-2}$.

\section{Metal Columns in the Post-Shock Cooling Layers}
\label{postcolumns}

In this section I present computations of the integrated metal column densities produced in
the post-shock cooling layers. In computing the columns, I integrate from the shock front to
the distance reached by the gas at the shock age. Older shocks extend over larger path-lengths
and therefore yield larger total columns. The complete shocks are integrated to a distance
where the gas has cooled down to the termination temperature, $T_{\rm low}=1000$~K.

The integrated metal ion column densities, $N_i$, through the post-shock cooling layers
are given by
\begin{equation}
\label{Ngeneral}
N_i = \int_0^{\rm age} n_{\rm H} Z A_{\rm el} x_i v dt\;,
\end{equation}
where $n_{\rm H}$ is the hydrogen density, $A_{\rm el}$ is the abundance of the element 
relative to hydrogen in solar metallicity gas, $X_i$ is the ion fraction, and $v$ is the
gas velocity.
The integration is over time, where
\begin{equation}
dt\propto \frac{dT}{n\Lambda(x_i,T,Z)}\;.
\end{equation}
The mass-continuity equation implies that $n_{\rm H}v$ is constant throughout the flow,
and the columns may therefore be expressed as
\begin{equation}
\label{Nprop}
n_i\propto Z \int_{T_{\rm age}}^{T_s} x_i \frac{dT}{n\Lambda}\;.
\end{equation}

As discussed in GS09, when $B=0$ the cooling times are shortened as the gas is
compressed during cooling, and the column densities of ions produced at low temperatures are 
significantly suppressed compared with strong-$B$ shocks.

\begin{figure*}[!h]
\epsscale{1.0}
\plotone{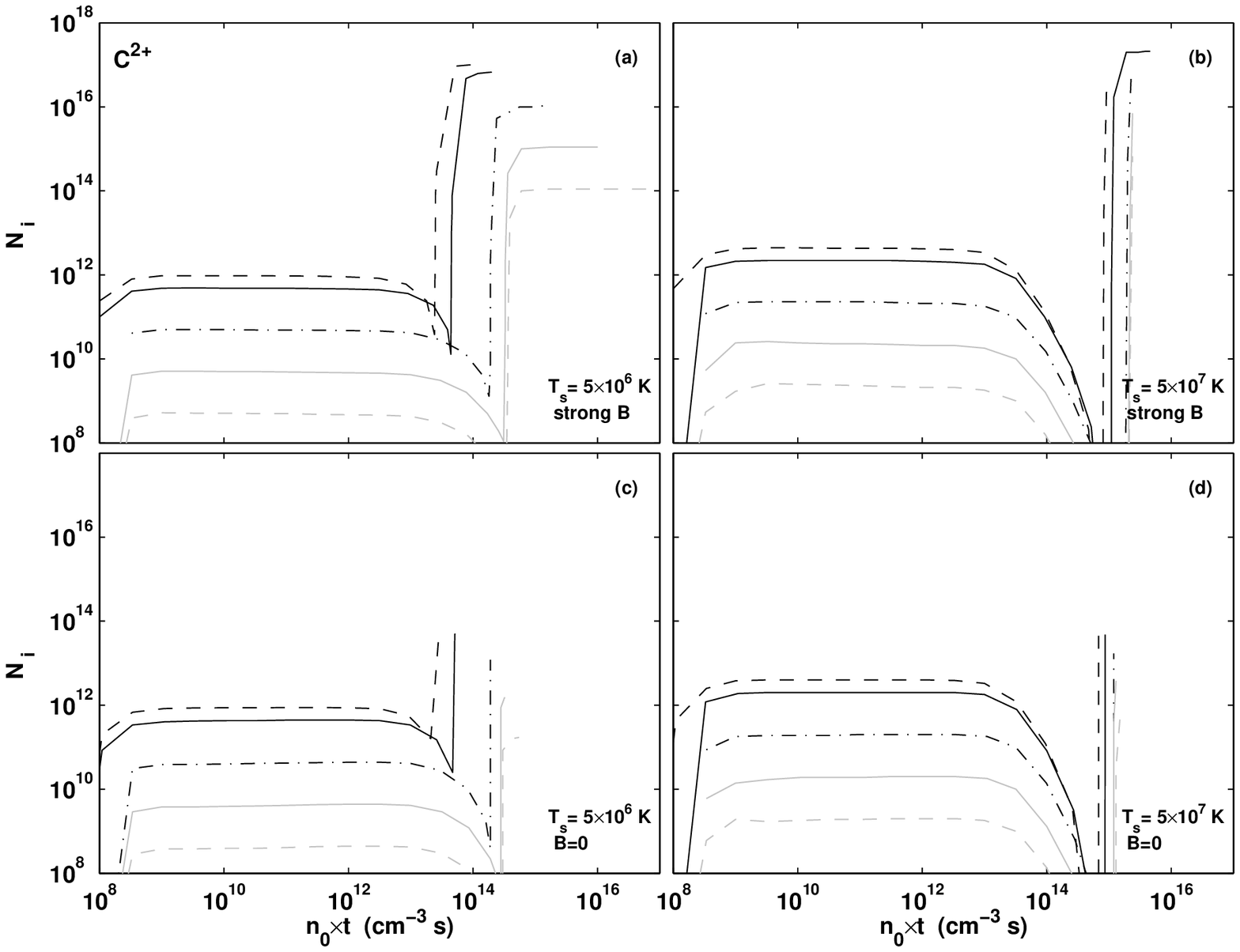}
\caption{Post-shock C$^{2+}$ columns as a function of shock age for $Z$ from $10^{-3}$ to $2$ times solar.
(a) $T_s=5\times10^6~K$, strong-$B$. (b) $T_s=5\times10^7~K$, strong-$B$. 
(c) $T_s=5\times10^6~K$, $B=0$. (d) $T_s=5\times10^7~K$, $B=0$.}
\label{post-CIII}
\end{figure*}

\begin{figure*}[!h]
\epsscale{1.0}
\plotone{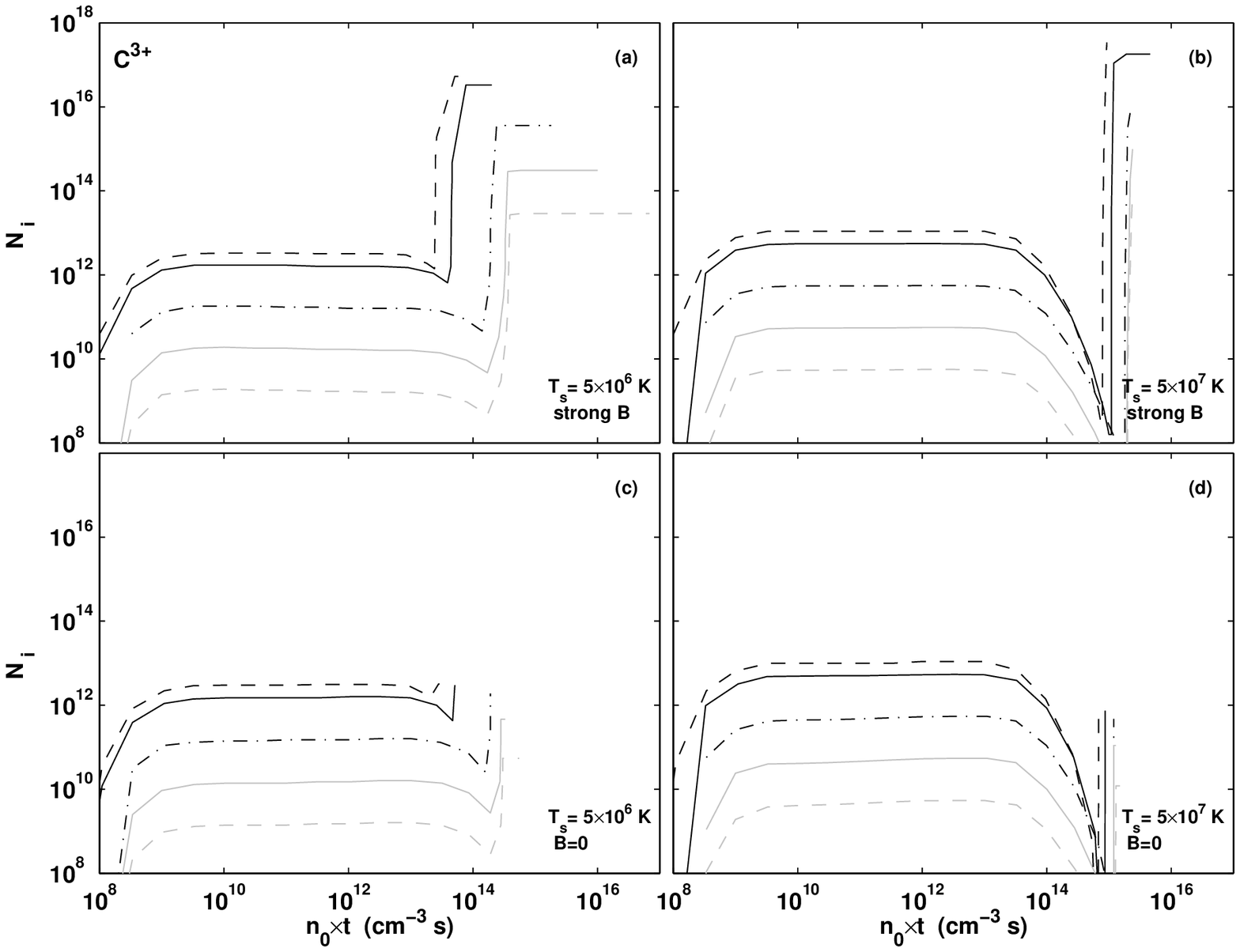}
\caption{Same as Figure~\ref{post-CIII}, but for C$^{3+}$.}
\label{post-CIV}
\end{figure*}

\begin{figure*}[!h]
\epsscale{1.0}
\plotone{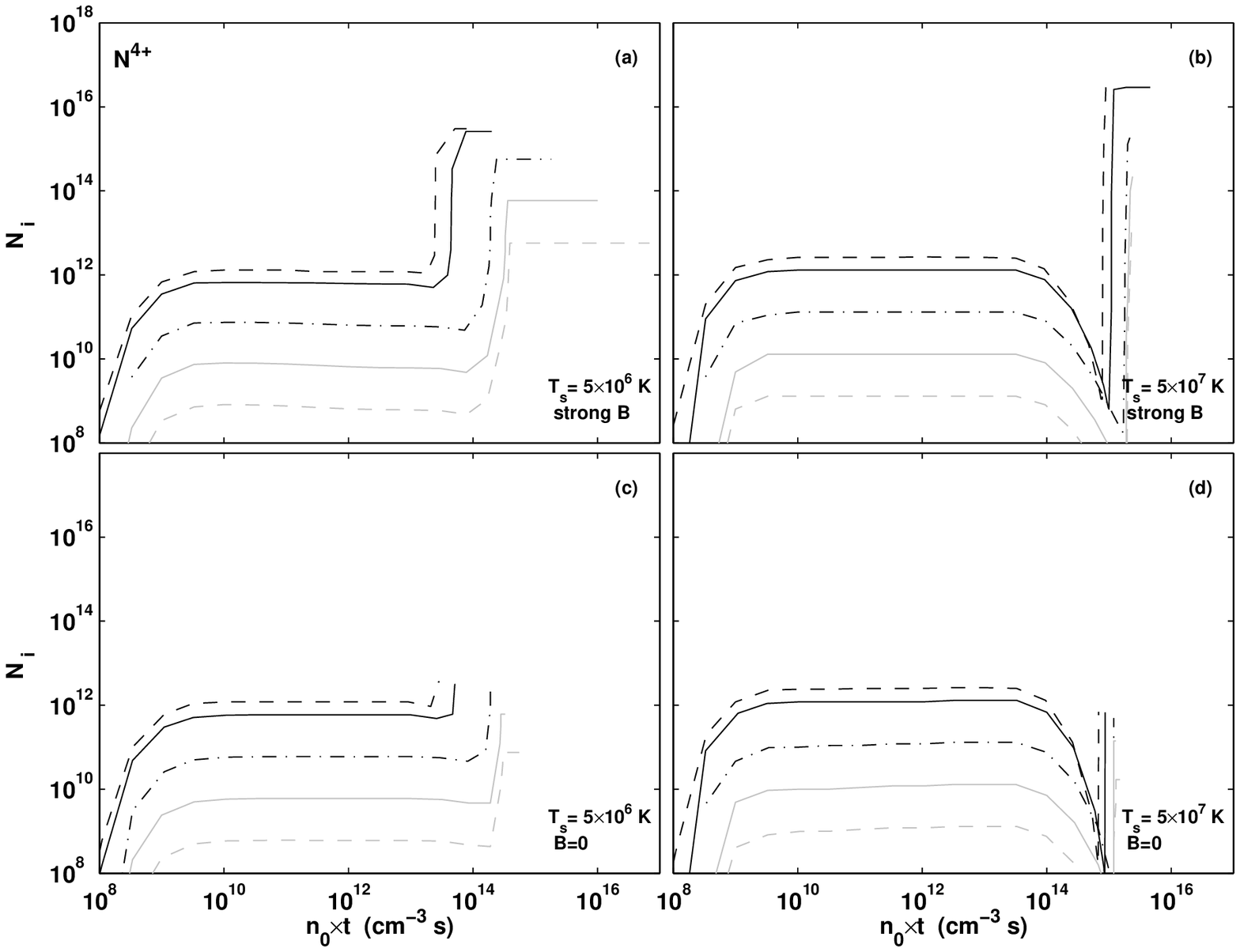}
\caption{Same as Figure~\ref{post-CIII}, but for N$^{4+}$.}
\label{post-NV}
\end{figure*}

\begin{figure*}[!h]
\epsscale{1.0}
\plotone{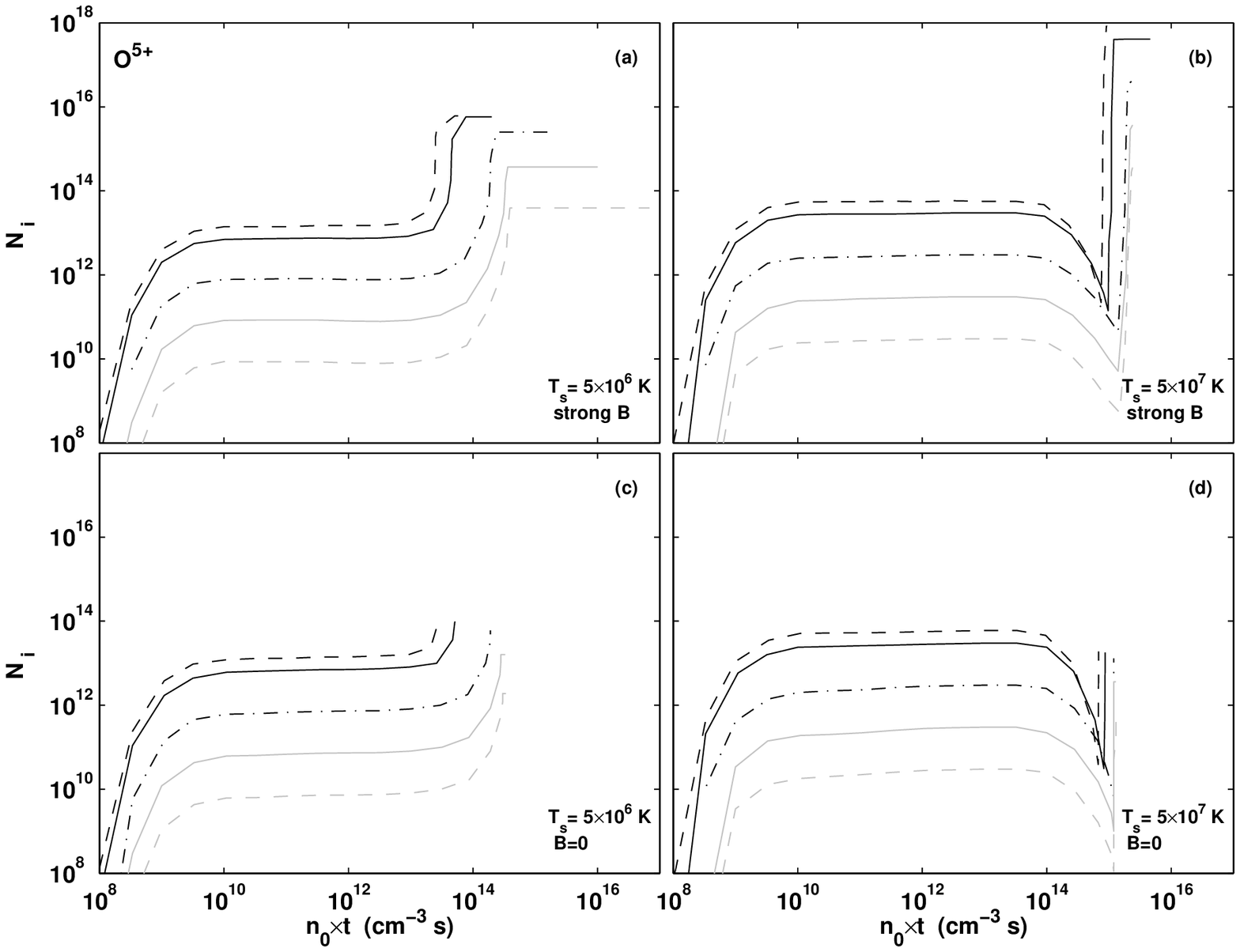}
\caption{Same as Figure~\ref{post-CIII}, but for O$^{5+}$.}
\label{post-OVI}
\end{figure*}

\begin{figure*}[!h]
\epsscale{1.0}
\plotone{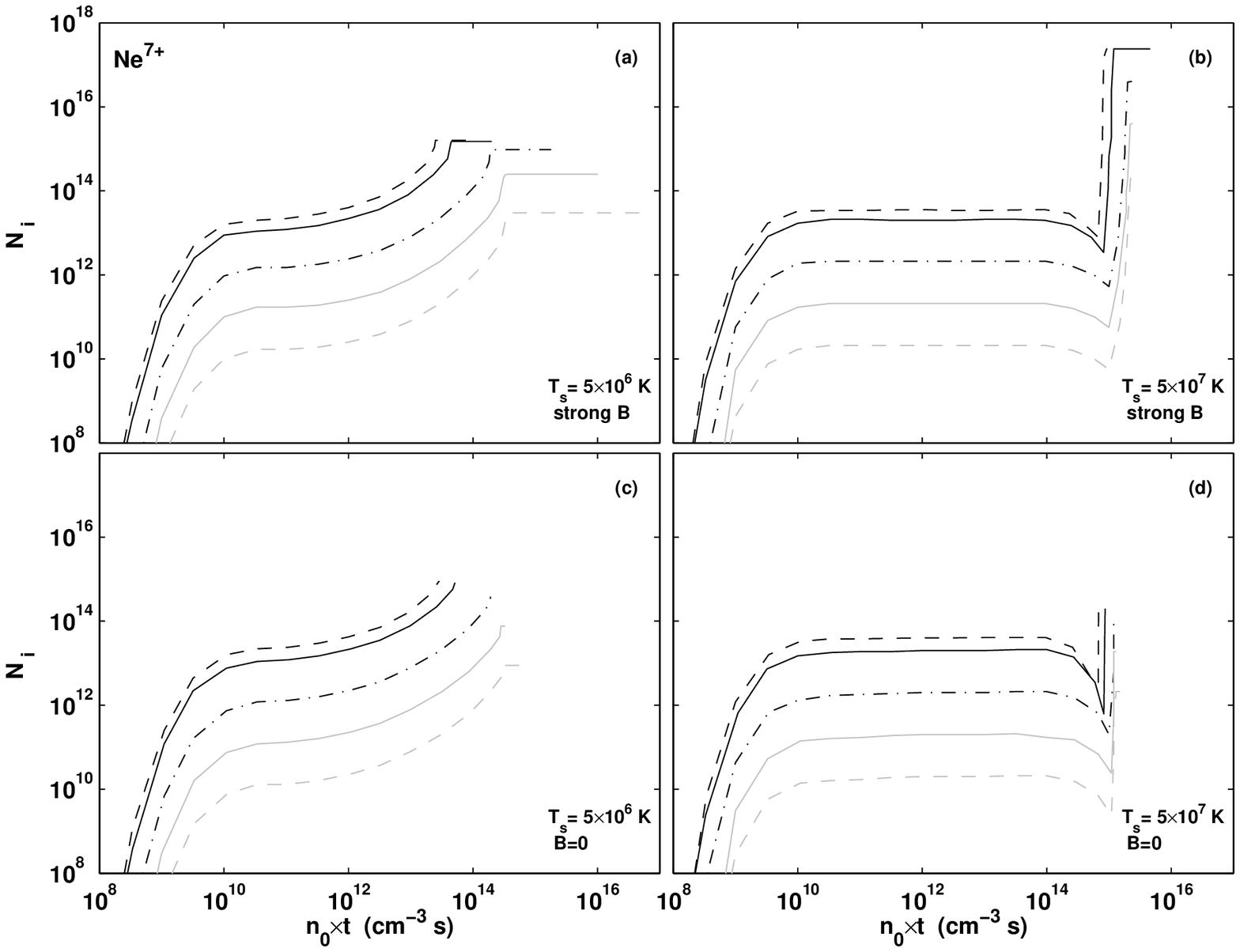}
\caption{Same as Figure~\ref{post-CIII}, but for Ne$^{7+}$.}
\label{post-NeVIII}
\end{figure*}

I list the full set of post-shock column densities as a function of shock
temperature, metallicity, magnetic field and age in Tables~\ref{post-table},
as outlined by Table~\ref{guide}.
 As examples, in Figures~\ref{post-CIII}-\ref{post-NeVIII}, I show the C$^{2+}$, C$^{3+}$,
N$^{4+}$, O$^{5+}$, and Ne$^{7+}$ post-shock column densities as a function of shock age,
for the various shock models considered here.

\begin{deluxetable}{lcccc}
\tablewidth{0pt}
\tablecaption{Post-Shock Columns vs. Age}
\tablehead{
\colhead{Age} &
\colhead{$N($H$^0)$} & 
\colhead{$N($H$^+)$} & 
\colhead{$N($He$^0)$} & 
\colhead{\ldots} \\
\colhead{(cm$^{-3}$~s)}&
\colhead{(cm$^{-2}$)}&
\colhead{(cm$^{-2}$)}&
\colhead{(cm$^{-2}$)}&
\colhead{(cm$^{-2}$)}}
\startdata
$1.000\times10^8$&$9.0\times10^{14}$&$6.7\times10^{14}$&$7.9\times10^{13}$ &\ldots\\
$3.300\times10^8$&$5.4\times10^{14}$&$4.4\times10^{15}$&$5.3\times10^{13}$ &\ldots\\
$1.000\times10^9$&$4.1\times10^{14}$&$1.5\times10^{16}$&$3.3\times10^{13}$ &\ldots\\
\enddata
\tablecomments{The complete version of this table is in 
the electronic edition of the Journal. The printed edition contains only a sample. 
The full table lists post-shock columns as functions of shock age for
the $B=0$ and strong-B limits,
for shock temperatures of $5\times10^6$~K and $5\times10^7$~K, and for
$Z=10^{-3}$, $10^{-2}$, $10^{-1}$, $1$, and $2$ times solar metallicity
gas (for a guide, see Table~\ref{guide}).}
\label{post-table}
\end{deluxetable}

\begin{deluxetable}{lllcc}
\tablewidth{0pt}
\tablecaption{Post-Shock and Precursor Columns Tables}
\tablehead{
\colhead{} & \colhead{} & \colhead{} &
 \colhead{Post-Shock} &\colhead{Precursor}\\
\colhead{Data} & \colhead{} & \colhead{} &
 \colhead{Columns} &\colhead{Columns}}
\startdata
$5\times10^6$~K & strong-$B$ & $Z=2$       & 2A & 4A \\
                &            & $Z=1$       & 2B & 4B \\
                &            & $Z=10^{-1}$ & 2C & 4C \\
                &            & $Z=10^{-2}$ & 2D & 4D \\
                &            & $Z=10^{-3}$ & 2E & 4E \\
                & $B=0$      & $Z=2$       & 2F & 4F \\
                &            & $Z=1$       & 2G & 4G \\
                &            & $Z=10^{-1}$ & 2H & 4H \\
                &            & $Z=10^{-2}$ & 2I & 4I \\
                &            & $Z=10^{-3}$ & 2J & 4J \\
$5\times10^7$~K & strong-$B$ & $Z=2$       & 2K & 4K \\
                &            & $Z=1$       & 2L & 4L \\
                &            & $Z=10^{-1}$ & 2M & 4M \\
                &            & $Z=10^{-2}$ & 2N & 4N \\
                &            & $Z=10^{-3}$ & 2O & 4O \\
                & $B=0$      & $Z=2$       & 2P & 4P \\
                &            & $Z=1$       & 2Q & 4Q \\
                &            & $Z=10^{-1}$ & 2R & 4R \\
                &            & $Z=10^{-2}$ & 2S & 4S \\
                &            & $Z=10^{-3}$ & 2T & 4T \\
\enddata
\label{guide}
\end{deluxetable}

Focus for example on Figure~\ref{post-CIV}. This figure confirms that the C$^{3+}$ column
is gradually built out of two components (see discussion in Section~\ref{example-columns}). 
Initially, C$^{3+}$ peaks as the gas is
ionized after passing the shock front. This first abundance peak is short, and rapidly
builds up a low column density, that then remains roughly constant with little additional
contribution as the shock ages because higher stages of ionization dominate the ion abundance. 
Finally, when the shock age is large enough to allow the gas to cool behind the shock front,
a second abundance peak is produced in the recombining layers. 
This second peak is much longer, and produces the 
significant column observed in the complete shocks (see Section~\ref{example-columns} for 
details). For example, panel~(a) shows that for $T_s=5\times10^6$~K and $Z=2$,
the C$^{3+}$ column that is produced during the initial ionization stage is 
$\sim10^{12}$~cm$^{-2}$.
This column is built up over $<10^9$~cm$^{-3}$~s, and then remains constant as higher
ionization states of carbon dominate in the hot radiative zone. Finally, when the shock age
reaches $\sim3\times10^{13}$~cm$^{-3}$~s, a second C$^{3+}$ is produced in the cooling layers.
This second dominant peak accumulates to a column density of $\sim5\times10^{16}$~cm$^{-2}$
for the complete shock.

The post-shock column densities are unaffected by the quasi-static approximation, because
the initial abundance peaks occur over a time scale shorter than the shock-age, and are
therefore insensitive to the details of the temperature profile, whereas the second dominant
abundance peaks are produces at late times within the nonequilibrium cooling zone and the
photoabsorption plateau, when the quasi-static approximation robustly holds.

Higher ionization states, that exist at or near the shock temperature are accumulated more
gradually over the shock age. For example, panel~(a) of Figure~\ref{post-NeVIII} shows that
in a $5\times10^6$~K shock, the Ne$^{7+}$ column density increases monotonically with time.
This is because Ne$^{7+}$ is abundant even within the hot radiative zone, and its column
density therefore accumulates more gradually. The Ne$^{7+}$ abundance peaks at a temperature
of $6\times10^5$~K, and indeed a rapid growth in its column can be seen at later times
as the shocked gas cools through this temperature.

\subsection{Gas Metallicity}

As discussed above, the gas metallicity affects the ion fractions produced in the 
post-shock cooling layers in several ways. First, higher metallicity gas cools more
efficiently. This results in shorter cooling times, and in increased recombination lags.
Second, the spectral energy distribution of the shock self-radiation is dominated by
line-produced UV-photons for high metallicity gas, and by a harder thermal bremsstrahlung
spectrum for lower-Z cases. Photoionization is therefore more efficient in higher-metallicity
shock, which produce larger ionization parameters (see also Figure~\ref{UvsZ}).

For the column density distributions shown in Figures~\ref{post-CIII}-\ref{post-NeVIII},
a distinction must be made between the first and second abundance peaks discussed above.
The first column peak is accumulated over an ionization time scale at the shock temperature.
This ionization time-scale does not depend on the gas metallicity, and is barely affected
by gas cooling, because little cooling occurs so rapidly.
The integrated metal ion column densities are therefore simply proportional to the gas 
metallicity (see Equation~\ref{Nprop}). This can clearly 
be seen in Figures~\ref{post-CIII}-\ref{post-NeVIII}.

The second abundance peak occurs during the post shock cooling layers.
The duration of this second peak is determined by the cooling time at the temperature
where the ion is produced. Equations~\ref{Nprop} shows that when the cooling function $\Lambda$
is proportional to $Z$ (when metals dominate the cooling) the columns are nearly independent
of gas metallicity. This is because when $\Lambda\propto Z$, reduced elemental abundances
are compensated by longer cooling times, and $Z$ cancels out in Equation~\ref{Nprop}.
The actual behavior is more complicated than that, because the ion distributions themselves
are functions of gas metallicity, due to departures from
equilibrium ionization, and because of the different spectral energy distributions of the
photoionizing self-radiation.

On the other hand, when the cooling efficiency is independent of $Z$ (because it is dominated
by hydrogen/helium cooling or by thermal bremsstrahlung), Equation~\ref{Nprop} indicates that
the metal-ion columns should be proportional to $Z$.

Figures~\ref{post-CIII}-\ref{post-NeVIII} confirm that for $B=0$ (when photoionization does
not significantly affect the ion distributions, see below) the final column densities are
only weakly dependent of metallicity when $Z\gtrsim0.1$ (i.e. metal dominate the cooling and
$\Lambda\propto Z$), and are roughly proportional to $Z$ for lower metallicities (when
H/He dominate the cooling).

\subsection{Shock temperature}

The shock temperature has a strong impact on the ion fractions and resulting column densities
produced in the post-shock cooling layers. First, faster shocks give rise to more highly
ionized species that are produced at or near the shock temperature. Second, as discussed
in Section~\ref{structure}, the energy flux radiated by the (complete) shock is proportional to
$v_s^3$ ($\propto T_s^{1.5}$). In addition, the spectral distribution of the shock self-radiation
is harder for faster shocks. Hotter shocks therefore produce an abundance of high energy 
photons, that can efficiently photoionize the gas in the cooling layers.

Equation~\ref{Ngeneral} implies that for ions that are produced near $T_s$, the metal columns
will be proportional to the shock velocity. This can be seen for the 
initial peaks shown in Figures~\ref{post-CIII}-\ref{post-NeVIII}.
The column densities produced in the initial peaks of the $5\times10^7$~K shocks
are a factor of $\sim3$ ($\sqrt{10}$) larger than those produced in the $5\times10^6$~K
shocks.
The complete column densities accumulated along the total cooling time of the shock are
affected by other factors, most importantly photoionization by the shock self-radiation.
This affects the ion distributions $x_i$ in Equation~\ref{Nprop}.

\subsection{Magnetic field}

The strength of the magnetic field is one of the parameters that determines the significance
of photoionization in the post-shock gas. In strong-$B$ shocks, the isochoric evolution 
ensures that the accumulated shock self-radiation propagates into constant density 
gas. However, when $B=0$ the gas is compressed as it cools, and the density in the photoabsorption
plateau is significantly larger than $n_0$. The ionization parameter is
accordingly lower, by a factor $T_s/T$. Photoionization is therefore very important in strong-$B$
shocks, but does not play a significant role when $B=0$.
In addition, the extent of the photoabsorption plateau is considerably shorter when $B=0$, due
to the faster cooling that takes place in the compressed gas.

As discussed above, Equation~\ref{Ngeneral} implies that the column densities of intermediate 
and low ion are significantly suppressed when $B=0$ compared to isochoric strong-$B$ limit.
This is confirmed by the final (complete shock) columns shown in 
Figures~\ref{post-CIII}-\ref{post-NeVIII}. For example, for a $5\times10^6$~K
shock in solar metallicity gas, a \ion{C}{4} column $>10^{16}$~cm$^{-2}$ is
created in the strong-$B$ limit, but this column is only $\sim10^{12}$~cm$^{-2}$ 
when $B=0$.
However, as discussed above, the initial evolution of the post-shock ion fraction 
following the shock front is not expected to depend on the intensity of the
magnetic field. Photoionization does not play an important role in the hot radiative
zone, and no significant compression takes place when $T\sim T_s$. The {\it initial}
columns are indeed very similar between the upper and lower panels of 
Figures~\ref{post-CIII}-\ref{post-NeVIII}.

\section{Metal Columns in the Radiative Precursor}
\label{precolumns}

In this section I present the equilibrium photoionization column densities that are
produced in the radiative precursors. The precursor gas is heated and photoionized 
by the shock self-radiation propagating upstream. This upstream radiation is gradually
being absorbed by the precursor gas, producing a photoabsorption layer in which the
heating and ionization rates gradually decline with increasing (upstream) distance from
the shock front.

For optically thin shocks, I use the post-shock flux at the position associated with
the shock age, $l_{\rm age}$, for the upstream radiation field.
For older shocks that become thick to their self-radiation, I use the post-shock flux
at the position where the gas first becomes thick at the Lyman limit, $l_{\rm thick}$.
Downstream of $l_{\rm thick}$, the radiation field is attenuated by absorption in the
post-shock photoabsorption plateau. 

For each shock model, I construct an equilibrium photoionization model assuming that 
the appropriate flux enters a gas of uniform density $n_0/4$.
The typical temperatures in the radiative precursor are in the range $10^4 - 10^5$~K,
and I integrate from the shock front to a distance where absorption reduces the heating 
rate sufficiently for the gas to reach a temperature $T_{\rm low} = 1000$~K.

To compute the photoionization ion fractions in the radiative precursors, I use the
photoionization code Cloudy (ver. 07.02.00) to construct equilibrium models of the precursor 
absorbing layers.
The ionization parameter in the radiative precursors are larger than in the corresponding
downstream photoabsorption plateaus, because the gas density in the
precursor is four times lower. Intermediate and low ions may therefore be produced more
efficiently in the precursor compared to the post-shock cooling layers, and the precursor
may dominate the total observed columns of such ions even for complete shocks (GS09).
For partial shocks, the column density of intermediate ions in the radiative precursor
builds up even before the post-shock gas cools to produce the photoabsorption plateus.
In such cases, UV absorption line signatures may be produced by the radiative precursor
while the shock gas is still too hot to be efficienctly observed.

\begin{figure*}[!h]
\epsscale{1.0}
\plotone{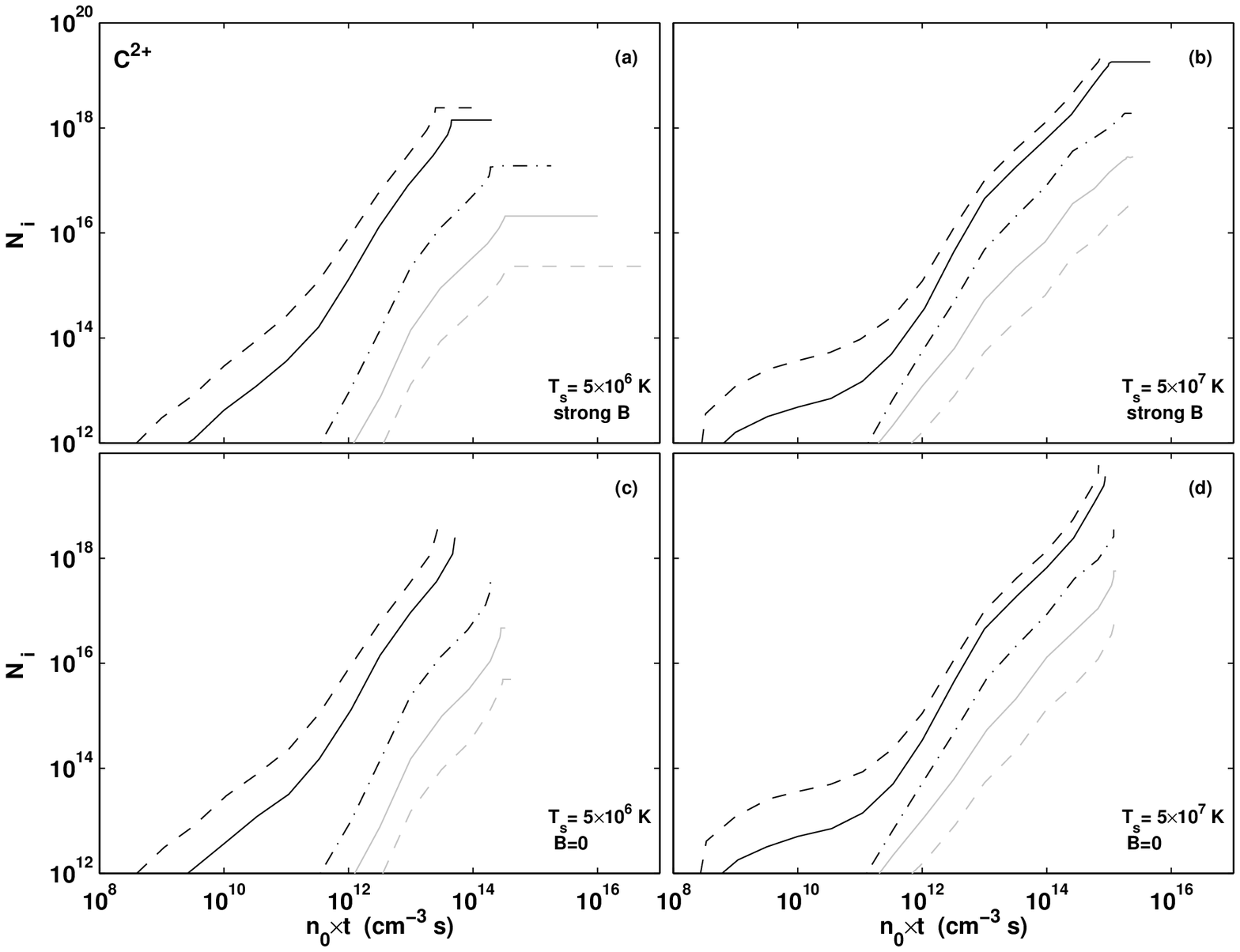}
\caption{Precursor C$^{2+}$ columns as a function of 
shock age for $Z$ from $10^{-3}$ to $2$ times solar.
(a) $T_s=5\times10^6~K$, strong-$B$. (b) $T_s=5\times10^7~K$, strong-$B$. 
(c) $T_s=5\times10^6~K$, $B=0$. (d) $T_s=5\times10^7~K$, $B=0$.}
\label{pre-CIII}
\end{figure*}

\begin{figure*}[!h]
\epsscale{1.0}
\plotone{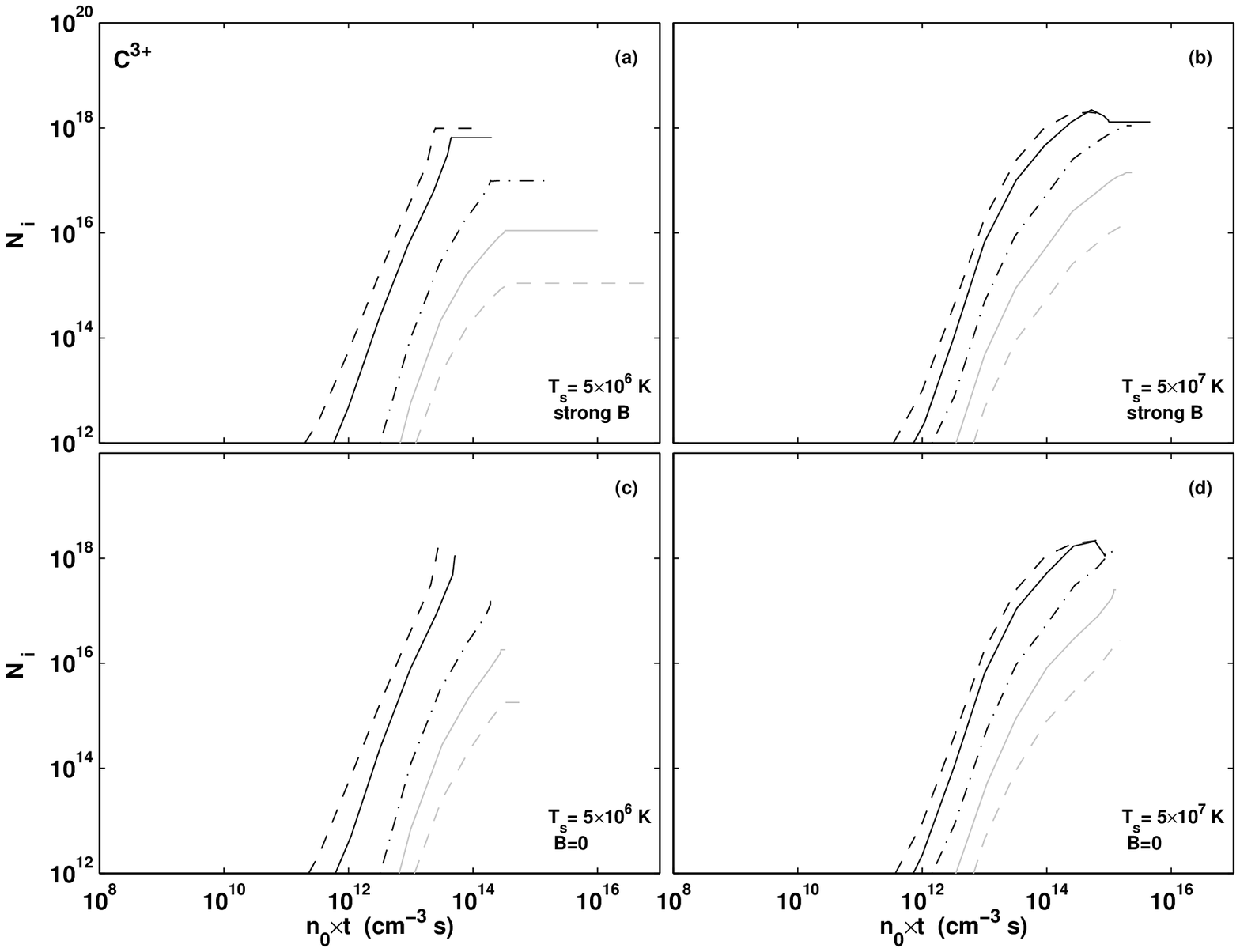}
\caption{Same as Figure~\ref{pre-CIII}, but for C$^{3+}$.}
\label{pre-CIV}
\end{figure*}

\begin{figure*}[!h]
\epsscale{1.0}
\plotone{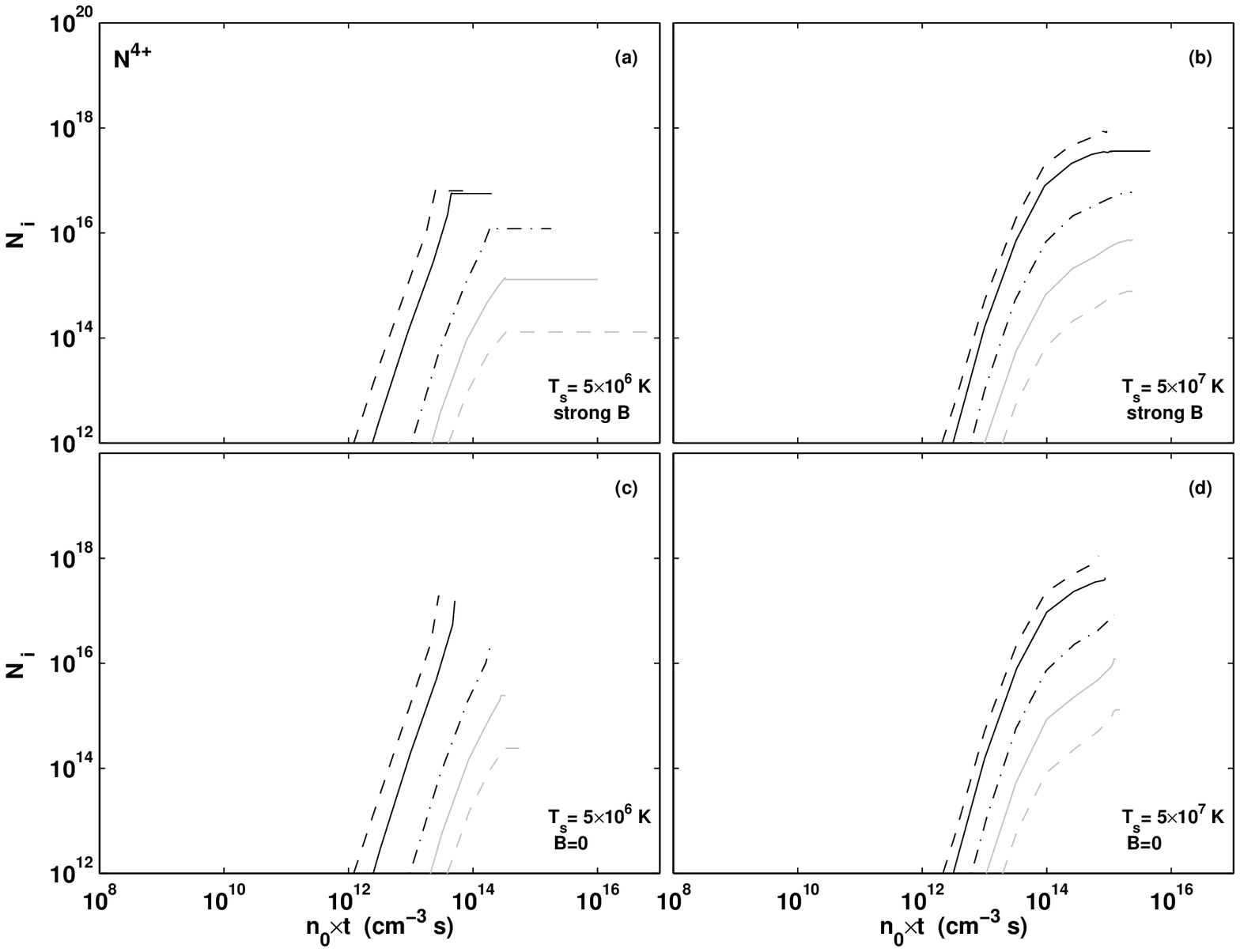}
\caption{Same as Figure~\ref{pre-CIII}, but for N$^{4+}$.}
\label{pre-NV}
\end{figure*}

\begin{figure*}[!h]
\epsscale{1.0}
\plotone{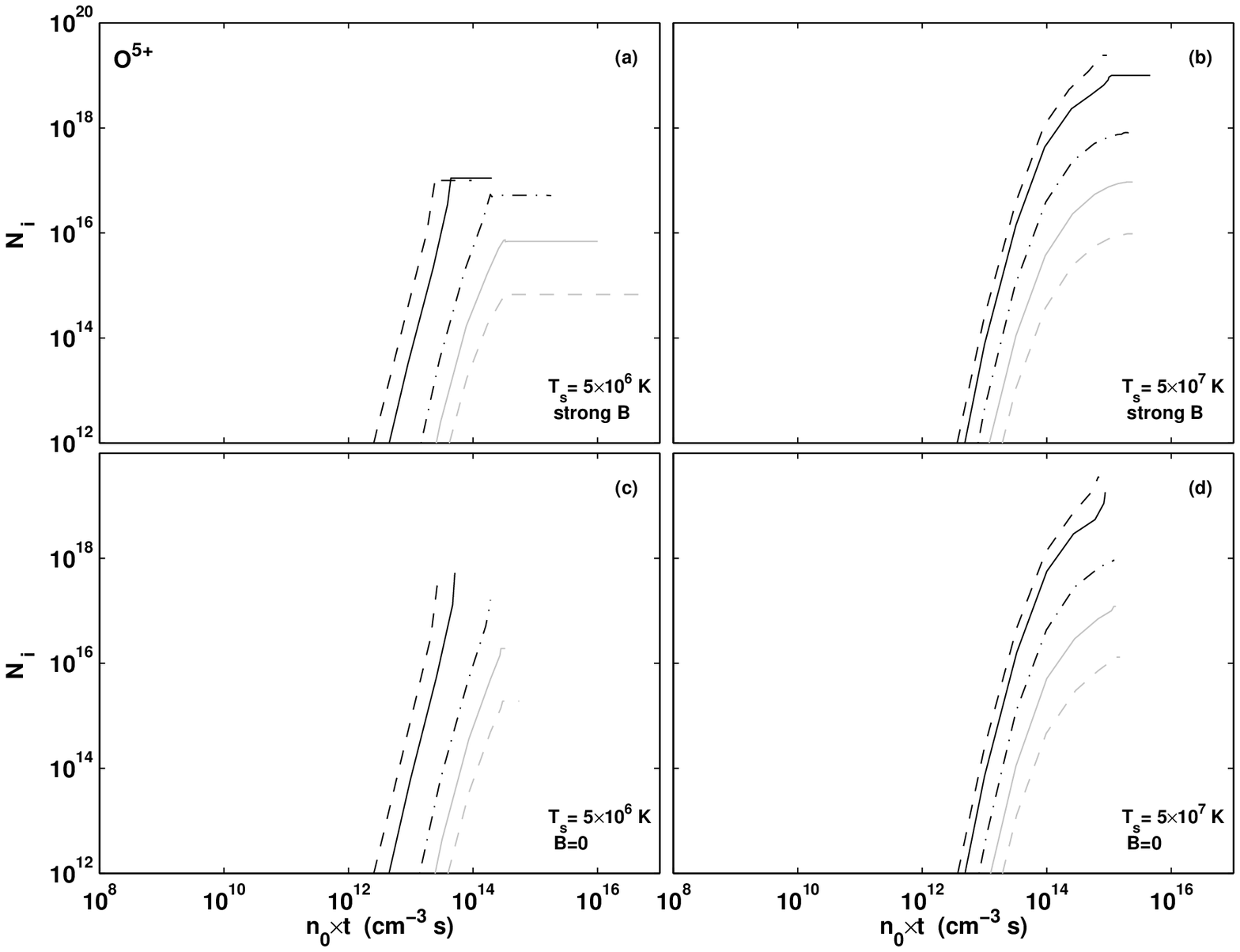}
\caption{Same as Figure~\ref{pre-CIII}, but for O$^{5+}$.}
\label{pre-OVI}
\end{figure*}

\begin{figure*}[!h]
\epsscale{1.0}
\plotone{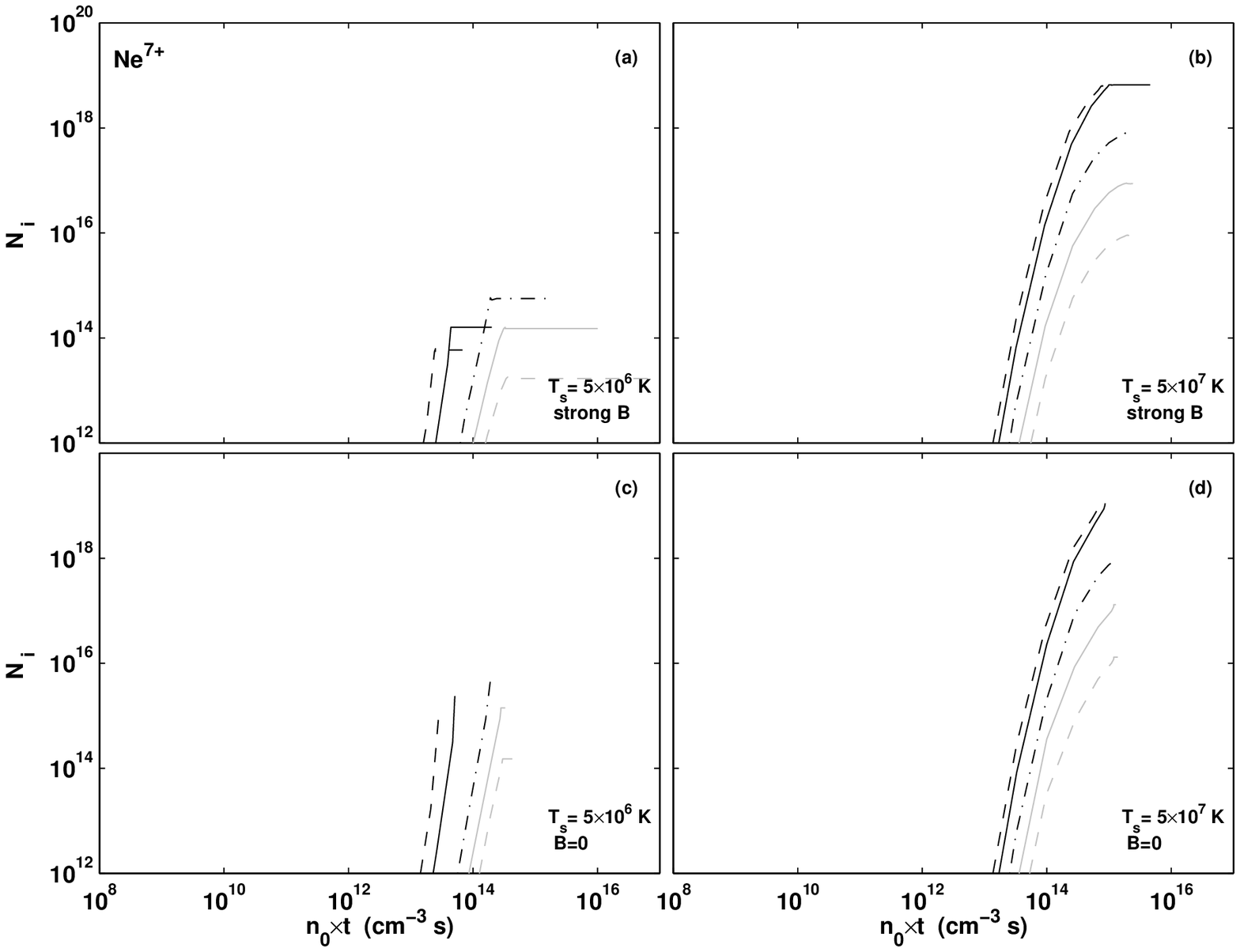}
\caption{Same as Figure~\ref{pre-CIII}, but for Ne$^{7+}$.}
\label{pre-NeVIII}
\end{figure*}

Figures~\ref{pre-CIII}-\ref{pre-NeVIII} display, as examples, the integrated precursor column
densities of C$^{2+}$, C$^{3+}$, N$^{4+}$, O$^{5+}$, and Ne$^{7+}$ as functions of the
shock age (cm$^{-3}$~s), for $Z$ between $10^{-3}$ and $2$. The upper panels are for
radiative precursors of strong-$B$ shocks, and the lower panels are for $B=0$.
The left hand panels are for $T_s=5\times10^6$~K, and the right hand panels for
$T_s=5\times10^7$~K. The full set of precursor column densities for all the metal ions that I 
consider are listed in Table~\ref{pre-table} as outlined in Table~\ref{guide}.

Figures~\ref{pre-CIII}-\ref{pre-NeVIII} confirm that as the flux entering the
precursor accumulates with time, deeper precursors are produced. As expected,
the precursor  columns increase monotonically with shock age.

\begin{deluxetable}{lcccc}
\tablewidth{0pt}
\tablecaption{Precursor Columns vs. Age}
\tablehead{
\colhead{Age} &
\colhead{$N($H$^0)$} & 
\colhead{$N($H$^+)$} & 
\colhead{$N($He$^0)$} & 
\colhead{\ldots} \\
\colhead{(cm$^{-3}$~s)}&
\colhead{(cm$^{-2}$)}&
\colhead{(cm$^{-2}$)}&
\colhead{(cm$^{-2}$)}&
\colhead{(cm$^{-2}$)} }
\startdata
$1.000\times10^8$&$1.6\times10^{17}$&$2.3\times10^{16}$&$1.4\times10^{16}$&\ldots\\
$3.300\times10^8$&$2.2\times10^{19}$&$2.7\times10^{18}$&$1.5\times10^{18}$&\ldots\\
$1.000\times10^9$&$3.6\times10^{19}$&$5.4\times10^{18}$&$2.1\times10^{18}$&\ldots\\
\enddata
\tablecomments{The complete version of this table is in 
the electronic edition of the Journal. The printed edition contains only a sample. 
The full table lists precursor columns as functions of shock age for
the $B=0$ and strong-B limits,
for shock temperatures of $5\times10^6$~K and $5\times10^7$~K, and for
$Z=10^{-3}$, $10^{-2}$, $10^{-1}$, $1$, and $2$ times solar metallicity
gas (for a guide, see Table~\ref{guide}).}
\label{pre-table}
\end{deluxetable}

Figures~\ref{pre-CIII}-\ref{pre-NeVIII} demonstrate that significant columns
of low (e.g.~C$^{2+}$) and intermediate/high (e.g.~C$^{3+}$, N$^{4+}$, O$^{5+}$) ions
are created in the radiative precursors.
For example, a comparison of Figures~\ref{pre-CIII} and~\ref{post-CIII} (note
different scales) shows that the C$^{2+}$ columns are dominated by the radiative 
precursors at all times.
This is also the case for C$^{3+}$ and N$^{4+}$ for $n_0t\gtrsim10^{12}$~cm$^{-3}$~s,
and for O$^{5+}$ at $n_0t\gtrsim10^{13}$~cm$^{-3}$~s.
However, for high-ions, which abundance peaks closer to $T_s$, the column densities
are dominated by the post-shock cooling layers. For example, the Ne$^{7+}$ CIE fractional 
abundance peaks at $\sim6\times10^5$~K, but it is abundant between $4\times10^5$ and 
$4\times10^6$~K (see Gnat \& Sternberg~2007). Figures~\ref{pre-NeVIII} and~\ref{post-NeVIII}
confirm that for $T_s=5\times10^6$~K, the \ion{Ne}{8} column produced in the post-shock
cooling layers dominates over the precursor column. This is because the equilibrium 
photoionization and heating rates in the precursor are too low to allow for the
efficient production of Ne$^{7+}$. However, for hotter $T_s=5\times10^7$~K shocks
(for which $T_s\gg T_{{\rm Ne}^{7+}}$), the \ion{Ne}{8} column is again dominated by the 
radiative precursor, as the more intense and energetic radiation field produced by the
hotter shocks efficiently ionize the precursor gas.

The column densities produced in the radiative precursors for strong-$B$ shocks are 
similar to those produced in $B=0$ precursors. The radiative fluxes in the two cases 
are similar, and differ only by a factor $5/3$ due to the $PdV$ work included in the
$B=0$ shocks.  A comparison of the upper and lower panels of 
Figures~\ref{pre-CIII}-\ref{pre-NeVIII} indeed shows that the columns are similar for all 
ages.

The ratio between the precursor and post-shock columns
is a strong function of shock age. Consider, for example, the C$^{3+}$ columns shown
in Figures~\ref{post-CIV} and~\ref{pre-CIV}. Figure~\ref{post-CIV}a shows that readily
detectable amounts of \ion{C}{4} ($\gtrsim10^{13}$~cm$^{-2}$) are only produced
in the post-shock cooling layers in the very final stages of the shock formation,
as the shocked gas cools through the non-equilibrium cooling zone and photoabsorption plateau.
For solar metallicity gas this occurs at $n_0t\sim5\times10^{13}$~cm$^{-3}$~s.
However, in the radiative precursor detectable amounts of \ion{C}{4} are produced as
early as $10^{12}$~cm$^{-3}$~s. This implies that there is a significant duration
of time over which fast radiative shocks may be observed by means of their radiative
precursor, even though the shocked gas itself was not able to cool, and therefore
produces little observable signatures.
A similar  behavior is also observed for other gas metallicities, temperatures,
and magnetic field properties.

I conclude that intermediate and high ions produced in the radiative precursors of
fast shock waves may serve as means for identifying and detecting young shocks that
have not yet existed over a cooling time, and  for which the shocked gas remains hot
and difficult to observe.

\section{Diagnostics}
\label{diagnostics}

\begin{figure}[!h]
\epsscale{1.18}
\plotone{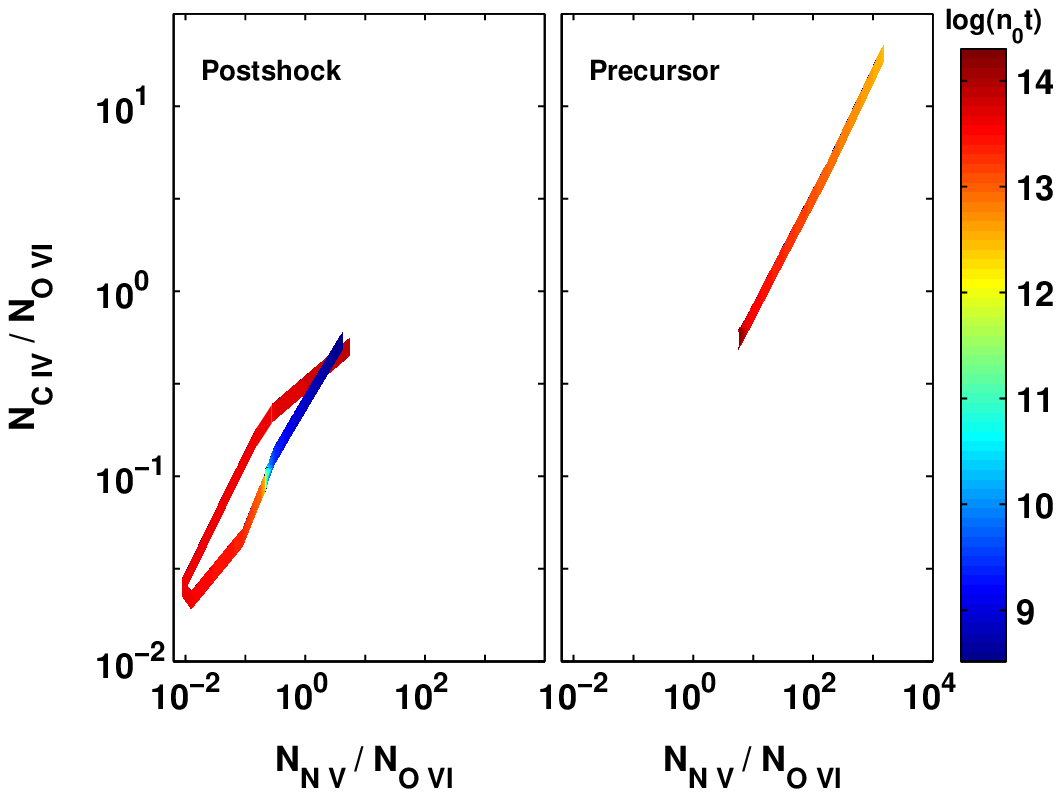}
\caption{Column density ratios $N_{{\rm C IV}}/N_{{\rm O VI}}$ vs. $N_{{\rm N V}}/N_{{\rm O VI}}$ 
is strong-$B$
shocks with $T=5\times10^6$ and $Z=1$~solar. Shock age is indicated by color along the 
trajectories, from young (blue) to old (red). Age (cm$^{-3}$~s) vs. color legend is on the right.
The left panel is for the shocked gas, and the right panel is for the radiative precursor.
The trajectories are only shown at ages for which the \ion{O}{6} column is greater than 
$10^{11}$~cm$^{-2}$.}
\label{eg-ratio}
\end{figure}

Diagnostic diagrams for fast radiative shocks may be constructed using the 
computational data presented in Sections~\ref{postcolumns} and~\ref{precolumns}. 
In Figure~\ref{eg-ratio} I show, as an example, ``trajectories'' for 
$N_{{\rm C IV}}/N_{{\rm O VI}}$ versus $N_{{\rm N V}}/N_{{\rm O VI}}$ for ages ranging from
$\sim3\times10^8$ to $\sim2\times10^{14}$~cm$^{-3}$~s, assuming a $5\times10^6$~K, $Z=1$,
strong-$B$ shock. The shock age is represented by color along the curves from young (blue)
to old (red). The trajectories are only shown at ages for which the absolute \ion{O}{6} 
column density is greater than $10^{11}$~cm$^{-2}$. The left panel is for the columns
produced in the post-shock cooling layers, and the right panel is for the radiative precursor.
Figure~\ref{eg-ratio} shows that the ion ratios created in the radiative precursor may be 
very different from those which are produced in the post-shock gas.
At early times ($n_0t\lesssim 5\times10^{13}$~cm$^{-3}$~s), the column densities
presented in Figure~\ref{eg-ratio} are completely dominated by the precursor columns.
At later times, the post-shock cooling layers also contribute to the observed columns.

As demonstrated above, the UV-absorption line signatures are dominated by
photoionized gas in the radiative precursor over a significant fraction
of the shock lifetime.
Ion-ratio diagrams, like the one shown in Figure~\ref{eg-ratio}, may serve as
diagnostics for gas in the radiative precursors of incomplete, fast radiative shocks.
The detection and identification of precursor gas may allow us to confirm the existence
of hot, unobservable, shocked gas, and to infer the associated baryonic content.

\section{Summary}
\label{summary}

In this paper, I present theoretical computations of the metal-ion column densities
produced in the post-shock cooling layers, and in the radiative precursors of
fast astrophysical shocks as a function of shock age.
My shock models rely upon the code and methods presented in Gnat \& 
Sternberg~(2009; GS09), but relax the assumption of steady-state complete shocks. 
To attain steady-state, the shocks must exist over time-scales that are long 
compared with their cooling times. Otherwise the shock structure, emitted 
radiation and radiative precursors are a function of the shock age in the 
partially cooled shocks.

I approximate the evolution of the shock properties by assuming quasi-static
evolution. For a given shock, as specified by the shock velocity, gas metallicity
and magnetic field, I construct a series of models, each appropriate for a specific
age, which are assumed to be independent of each other.

In these models, I use and extend the shock code developed in GS09. This code computes
the time-dependent downstream ion fractions and cooling efficiencies, 
follows the radiative transfer of the shock self-radiation through the 
post-shock cooling layers, takes into account the resulting photoionization and heating 
rates, follows the one-dimensional dynamics of the flowing gas, and self-consistently
computes the photoionization ion fractions in the radiative precursor.
I follow the ion fractions of all ionization states of the elements H, He, C, N, O,
Ne, Mg, Si, S, and Fe. I consider shock velocities of $600$ and $2000$~km~s$^{-1}$,
associated with shock temperatures of $5\times10^6$ and $5\times10^7$~K,
and gas metallicities between $10^{-3}$ and $2$ times the heavy element
abundance of the Sun.

For the dynamical evolution I consider shocks with no magnetic field ($B=0$),
for which the gas evolves with nearly constant pressure, with $P_\infty=4/3\;P_0$,
where $P_0$ is the post-shock pressure. I also consider shocks in which the 
magnetic field is dynamically dominant everywhere ($B/\sqrt{\rho}\gg v_s$)
so that the pressure is dominated by the magnetic field, and the density
remains constant in the flow. When $B=0$ the gas is compressed and decelerated
as is cools. The increased density implies faster cooling, and the low-temperature
evolution is significantly faster when $B=0$ compared with the strong-$B$ limit.
In Section~\ref{physics}, I describe the equations that I solve,
and the numerical method used to compute the shock properties.

In Section~\ref{structure}, I investigate the shock structure, and focus on how 
it depends
on the shock age. Complete shocks are composed of several zones, through which
the gas flows. The precursor gas enters the hot radiative zone, in which
the gas is hot and cooling is inefficient. After a significant fraction of
the shock input energy is radiated, the temperature finally declines, allowing
the cooling rate to increase. The gas then goes through the nonequilibrium cooling
zone, in which cooling is very efficient, and recombination lags may occur.
Once the neutral fraction rises, photoabsorption becomes efficient, and the gas
enters a photoabsorption plateau in which the temperature profile is shallower,
until finally all self-radiation is absorbed and the gas cools completely.
In partial, time-dependent shocks, this classical picture is modified due to the
finite shock age. The shock structure, self-radiation and precursor ionization
then depend on the shock age.

The shock self-radiation accumulates over time, as deeper downstream layers
contribute to the emitted radiation.
Initially, the self-radiation of young shocks is faint, leading to
low ionization parameters and large neutral fractions in their radiative 
precursors. In this case, the post-shock free electron abundance
is small, leading to inefficient excitation and ionization in the downstream
gas. As the intensity of the self-radiation increases, the neutral fraction in the
radiative precursor diminishes. At this stage, both free electrons
and neutral hydrogen atoms are abundant in the gas entering the hot radiative zone.
Cooling due to both collisional ionization and Ly$\alpha$ emission is highly
efficient under these conditions. Finally, once the shock self-radiation is
intense enough to fully ionize the precursor hydrogen gas, the initial cooling
drops again due to the lack of efficient coolants. Even then, cooling remains 
large compared with CIE, due to the underionized metal species that exist in 
the radiative precursor.

As expected, the ionization parameter in the radiative precursor increases
monotonically with shock age as the shock self-radiation builds up. 
After the shock exists for a cooling time, the intensity of this radiation 
stabilizes at a value appropriate for the complete shock.
I find that the precursor ionization parameter is a strong function of gas
metallicity. For low metallicities, the shock self-radiation is composed
almost entirely of thermal bremsstrahlung emission. The precursor ionization
parameter is then a simple power law in the shock age.
However, for larger metallicities, a significant fractions of the shock
self-radiation is emitted as metal-lines, creating a ``UV-bump''
in the spectral energy distribution. The contribution of the metal-line
emission enhances the ionization parameter, which may be between
$0.5$ and $2$ dex larger than the bremsstrahlung value.
Hotter shocks emit a larger fraction of their input energy as thermal 
bremsstrahlung even at high metallicities, and their ionization parameters
are closer to the free-free values.

In Sections~\ref{example-columns}-\ref{postcolumns}, I present computations of
the integrated metal-ion column densities in the post-shock cooling layers.
I examine how the downstream column densities evolve
with time. I list the complete set of post-shock column densities
in Table~\ref{post-table}, as outlined in Table~\ref{guide}.
For UV-detectable ions (\ion{C}{4}, \ion{N}{5}, \ion{O}{6} etc.),
the column densities are composed of two distinct contributions. First, some 
column density is initially created over a very short time scale (an ionization
time-scale), as the underionized gas in the precursor approaches CIE at the 
shock temperature. This column density builds over a time scale which is orders
of magnitudes shorter than the cooling time, and later remains constant as the 
shock evolves. For example, for \ion{C}{4} the initial column density in a 
$5\times10^6$~K, $Z=1$, strong-$B$ shock is $\sim10^{12}$~cm$^{-2}$. It accumulates
over a time scale of $\sim10^9$~cm$^{-3}$~s, and later remains constant for
a few $10^{13}$~cm$^{-3}$~s.
A second, dominant peak is later created as the shocked gas cools and recombines
through the nonequilibrium cooling zone and the photoabsorption plateau.
In the example above, the second peak yields a \ion{C}{4} column density of
$\sim3\times10^{16}$~cm$^{-2}$, which is the column observed in a complete
shock.

The initial peaks produce very low column densities which are difficult to observe.
The columns in these initial peaks are roughly proportional to gas metallicity.
The second dominant peak depends on the gas metallicity in a more complicated
manner, related to the dominant coolants at various temperatures.
The initial peaks are also insensitive to the value of the magnetic field, as
they are created on an ionization time-scale over which very little
evolution in the thermal and dynamical properties of the gas takes place.
However, the second dominant peaks are strongly affected by the magnetic field.
When $B=0$, the compression and associated accelerated cooling lead to significant
suppression of intermediate and low ions created at low temperatures compared
with constant-density, strong-$B$ shocks. High-ions, which are created at
or near the shock temperature, do not depend on the value of the magnetic field.
UV observations of shocked gas, and in particular the ratio of high to intermediate
ions, may be used to probe the intensity of interstellar and intergalactic magnetic
fields in complete fast shocks. 

In Section~\ref{precolumns}, I compute the precursor column densities for the 
various shock models as functions of shock age. As expected, the column densities 
of the intermediate and high ions are monotonically increasing functions of the 
shock age. They grow as the intensity of the shock self-radiation accumulates with 
time, leading to more highly-ionized, deeper upstream absorption layers.
I list the full set of precursor column densities in Table~\ref{pre-table}, as 
outlined in Table~\ref{guide}.
I find that the precursor column densities of intermediate and low ions dominate
the observable signatures from fast shocks over a significant fraction of the
shock cooling time. 

As opposed to the column densities created in the downstream cooling layers,
the precursor column densities are insensitive to the value of the magnetic field.
Because the shock self-radiation is dominated by emission in the hot radiative zone,
in which no significant cooling takes place, the ionization parameter and spectral
energy distributions are similar for strong-$B$ shocks and for shock in which $B=0$.

The results presented here demonstrate that over extended periods of time, the
precursor produces observable amounts of intermediate and high ions, such as
\ion{C}{4}, \ion{N}{5}, and \ion{O}{6}. These precursor UV signatures may be 
the only means to detect and identify the existence of young fast shocks.
For example, in a $5\times10^6$~K, solar metallicity shock in the strong-$B$ limit,
readily detectable amounts of \ion{C}{4} ($\gtrsim10^{13}$~cm$^{-2}$) are only produced
in the post-shock cooling layers at very late times, at ages $\gtrsim5\times 
10^{13}$~cm$^{-3}$~s. However, in the radiative precursor similar amounts of 
\ion{C}{4} are produced as early as $10^{12}$~cm$^{-3}$~s.

Finally, in Section~\ref{diagnostics}, I demonstrate how ion-ratio diagrams may 
serve as diagnostics for gas in the radiative precursors of partially-cooled
fast astrophysical shocks. This provides a mean to observationally identify
the precursors of young fast shocks. I suggest that the detection and 
identification of precursor gas, photoionized by the hard shock self-radiation,
may allow us to confirm the existence of the hot, unobservable, shocked counterpart.
Absolute calibration of the observed precursor column densities may then be used
to infer the shocked-gas column, and estimate the "missing" baryonic mass 
associated with the shocks.

%

\section{Conclusions}
\label{conclusions}

Shock waves are a common phenomenon in astrophysics, and have a profound 
impact on the energetics and distribution of gas in the interstellar and
intergalactic medium. In this paper, I present theoretical computations of 
the metal-ion column densities produced in the post-shock cooling layers, 
and in the radiative precursors of fast astrophysical shocks as a function 
of shock age. 

This work is motivated by recent UV and X-ray observations
of gas that may be a part of a $10^5-10^7$~K ``warm/hot intergalactic medium''
(e.g.~Tripp et al.~2007; Savage et al.~2005; Buote et al.~2009; Narayanan 
et al.~2009; Fang et al.~2010; Danforth et al.~2010). However, the results
presented here are also applicable to other astrophysical environments in 
which young shocks are expected to prevail, including supernovae remnants 
and active galactic nuclei. The WHIM application is particularly interesting,
because the warm/hot intergalactic medium is expected to be a major reservoir 
of baryons (e.g.~Cen \& Ostriker~1999; Dav{\'e} et al.~2001; Bertone et 
al.~2008), and in fact account for the so-called ``missing baryons'' in 
the low-redshift Universe.

In my numerical models, the time-evolution of the shock structure, 
self-radiation, and associated metal-ion column densities are computed
by a series of quasi-static models, each appropriate for a different shock age.
The results for the post-shock and precursor columns are presented in convenient
online format in Tables~\ref{post-table} and~\ref{pre-table}.
For the post-shock columns, I use a shock code that explicitly follows the
nonequilibrium ionization states in the post-shock cooling layer, taking into 
account the impact of the shock self-radiation, cooling, and gas dynamics.
The precursor columns are set by photoionization equilibrium with the shock 
self-radiation, which depends on the shock parameters and age.

The shock models presented here are applicable to the shock-heated warm/hot
intergalactic medium, and may provide a way to identify and measure some
of the missing baryonic mass. My computations indicate that readily observable
amounts of intermediate and high ions, such as \ion{C}{4}, \ion{N}{5}, and
\ion{O}{6} are created in the precursors of young shocks, for which the
shocked gas remains hot and difficult to observe (as in the WHIM). 

I suggest that such precursors may provide a way to identify WHIM shocks: The 
absorption-line signatures predicted here may be used to construct ion-ratio
diagrams (e.g. Figure~\ref{eg-ratio}), which will serve as diagnostics
for the photoionized gas in the precursors. 
Absolute calibration of the observed precursor column densities -- using
the data listed in Tables ~\ref{post-table} and~\ref{pre-table} -- may 
then be used to infer the shocked-gas column, and estimate the "missing"
baryonic mass associated with the shocks.

\section*{Acknowledgments}

My research is supported by NASA through Chandra Postdoctoral
Fellowship grant number PF8-90053 awarded by the Chandra X-ray Center, 
which is operated by the Smithsonian Astrophysical Observatory for NASA 
under contract NAS8-0306.


\end{document}